\documentstyle[preprint,aps]{revtex}

\tightenlines

\begin{document}

\draft

\title{\bf Density Matrix Renormalisation Group Approach to the Massive 
Schwinger Model}
\author{T. Byrnes, P. Sriganesh, R.J. Bursill, and C.J. Hamer.
}
\address{School of Physics,                                              
The University of New South Wales,                                   
Sydney, NSW 2052, Australia.}                      

\date{\today}

\maketitle 

\begin{abstract}
The massive Schwinger model is studied, using a
density matrix renormalisation group approach to the staggered lattice
Hamiltonian version of the model. Lattice sizes up to 256 sites are
calculated, and the estimates in the continuum limit are almost two
orders of magnitude more accurate than previous calculations. Coleman's
picture of `half-asymptotic' particles at background field $\theta =
\pi$ is confirmed. The predicted
phase transition at finite fermion mass $ (m/g) $ is accurately located, and
demonstrated to belong in the 2D Ising universality class.
\end{abstract}                    

\narrowtext

\section{INTRODUCTION}
The Schwinger model\cite{schwinger62,lowenstein71}, or
quantum electrodynamics in one space and one time dimension, exhibits
many analogies with QCD, including
confinement, chiral symmetry
breaking, charge shielding, and a topological
$\theta$-vacuum\cite{coleman75,casher74,casher75,coleman76}. It is a common 
test bed for the trial of new techniques for the study of QCD:
for instance, several authors\cite{creutz95,narayanan95,gattringer96,kiskis00}
 have recently discussed new methods
of treating lattice fermions using the Schwinger model as an example.

	Our purpose in this paper is twofold. First, we aim to explore
the physics of this model when an external `background' electric field
is applied, as discussed long ago in a beautiful paper by
Coleman\cite{coleman76}. Secondly, we wish to demonstrate the
application of density matrix renormalisation group methods \cite{white92,gehring97} to a
model of this sort, with long-range, non-local Coulomb interactions.

	Coleman\cite{coleman76} showed that the physics of the Schwinger
model is periodic in $\theta = 2{\pi}F/g$, where $ F $ is the applied
`background' electric field, and $g$ is the elementary charge. In the
special case $\theta = \pi$, some amusing phenomena occur. Whereas the
``quarks" in this model are generally confined by a classical linear
potential, at $\theta = \pi$ it is possible for single quarks to appear
deconfined, i.e. move freely, provided they do not cross another quark
(``half-asymptotic particles"). Coleman\cite{coleman76} also
demonstrated that for $\theta =\pi$ a phase transition must occur at
some finite value of $m/g$, where $m$ is the quark mass. These arguments
are reviewed more fully in Section II.

	There have been very few attempts to verify these predictions 
numerically, that we are aware
of. Hamer, Kogut, Crewther and Mazzolini\cite{hamer82} 
used finite-lattice techniques to address the problem. They calculated the 
ground-state
energy and `string tension' as functions of $\theta$ and $m/g$. They
located the phase transition at $\theta = \pi$ to lie at $(m/g)_{c} =
0.325(20)$, with a correlation length index $\nu = 0.9(1)$. They also
attempted to estimate the chiral order parameter; but it was later
pointed out \cite{burden88} that the chiral order parameter actually
suffers from a logarithmic divergence at finite $m/g$ in this model.
Schiller and Ranft \cite{schiller82} used Monte Carlo techniques to
locate the phase transition at $(m/g)_{c} = 0.31(1)$.

	 Many different numerical methods have been applied to the
Schwinger model in zero background field, including
strong-coupling series \cite{banks76,carroll76,berruto98,hamer97}, finite-lattice
calculations\cite{crewther80,irving83,sriganesh00}, Monte Carlo
calculations\cite{martin82,schiller82,carson86,baillie87,azcoiti94}, 
discrete light-cone
quantization\cite{eller87} and the related light-front
Tamm-Dancoff{\cite{bergknoff77,mo93} and ``fast-moving frame"\cite{kroger98}
techniques, a recently proposed ``contractor renormalisation group"
method\cite{melnikov00}, and finally a coupled-cluster expansion\cite{fang01}. 
For a recent review, see Sriganesh {\it et al.} \cite{sriganesh00}.
	Analytic calculations of the mass spectrum have been carried out
using mass perturbation theory\cite{coleman75,vary96,adam96,harada95}
for small fermion masses, and weak-coupling
expansions\cite{coleman76,hamer77,sriganesh00} for large fermion masses.

	In this paper, we apply a new technique, namely the density
matrix renormalisation group (DMRG), which has been used with great
success\cite{white92,gehring97} for lattice spin models and lattice 
electron models such
as the Hubbard model. 
It was also recently applied to a simple one-particle potential model
with asymptotic freedom by Martin-Delgado and
Sierra\cite{martindelgado99}.
It might be questioned
whether the DMRG can successfully handle a model involving long-range
Coulomb interactions such as the Schwinger model. We perform some
calculations in zero background field, to show that in fact the approach
works extremely well. It gives accurate results for lattices of up to
256 sites, and provides estimates of the continuum limit which are
around 50 times more accurate than previous calculations.

	The layout of the paper is as follows. In Section II we review
Coleman's discussion \cite{coleman76} of the Schwinger model in a
background field, and the lattice formulation thereof. In Section III
the DMRG technique is outlined. In Section IV our main results at
background field $\theta = \pi$ are presented. Mass gaps and order
parameters are calculated, and the critical parameters at the phase
transition are accurately estimated, showing that the transition lies in
the universality class of the transverse Ising model in (1+1)
dimensions. The `half-asymptotic' particles behave
in exactly the same way as `kinks' or `spinons' in the transverse Ising model,
giving rise to some surprising effects. Finally, in Section V some
results at $\theta =0$ are presented, and compared with previous
results, both analytical and numerical. Our conclusions are summarized
in Section VI.

\section{THE SCHWINGER MODEL IN A BACKGROUND FIELD}

\subsection{Fermion formulation}

The Schwinger model Lagrangian density, in standard notation, is
\begin{equation}
{\cal L} = - {1\over 4} F_{\mu \nu} F^{\mu \nu} + 
\bar{\psi} (i \not\!\partial - g \not\!\!A - m) \psi
\end{equation}
where $ \psi $ is a 2-component spinor field, since there is no spin in one space dimension. The coupling $ g $ has dimensions of mass, so the theory is super-renormalisable. Using $ g $ as the scale of energy, the physical properties of the model are then functions of the dimensionless ratio $ m/g $. The field strength term is
\begin{equation}
F_{\mu \nu} = \partial_{\mu} A_{\nu} - \partial_{\nu} A_{\mu}.
\end{equation}
The equations of motion are the Dirac equation
\begin{equation}
( i \not\!\partial - g \not \! \! A -m) \psi = 0
\end{equation}
and Maxwell's equations
\begin{equation}
\partial_\mu F^{\mu \nu} = g \bar{\psi} \gamma^\nu \psi.
\end{equation}

Choosing a time-like axial gauge
\begin{equation}
A_0 = 0
\end{equation}
the field-strength tensor reduces to
\begin{equation}
F^{10}= - \dot{A}^1 = E
\end{equation}
where $ E $ is the 1-component electric field. Gauss' law becomes
\begin{equation}
\partial_1 E = - \partial_1 \dot{A}^1 = g j^0 = g \bar{\psi} \gamma^0 \psi.
\label{guass}
\end{equation}
The Hamiltonian becomes
\begin{equation} \label{fermionham}
 H = \int dx \left( - i \bar{\psi} \gamma^1 (\partial_1 + i g A_1 ) \psi + m \bar{\psi} \psi + {1\over 2} E^2 \right).
\end{equation}

Gauss' law can be integrated to give
\begin{equation}
E = g \int dx j^0(x) + F,
\end{equation}
showing that $ E $ is not an independent field, but can be determined in terms 
of the charge density $ j^0(x) $, up to the constant of integration $ F $, 
which corresponds to a ``background field'', as discussed by Coleman \cite{coleman76}. We can think of the background field as created by condenser plates at either end of our one-dimensional universe.

If $ |F| > g/2 $, charged pairs will be produced, and separate to infinity, until the field is reduced within the range $ |F| \leq g/2 $, thus lowering the electrostatic energy per unit length (Fig. \ref{fig:pair}). Thus physics is periodic in F with period $ g $, and it is convenient to define an angle $ \theta $ by
\begin{equation}
\theta = 2 \pi\frac{F}{g}.
\end{equation}
Then we can always choose $ \theta $ to lie in the interval $ [0,2 \pi ] $.

In the weak-coupling limit $ m/g \rightarrow \infty $, the vacuum contains no fermionic excitations, and the vacuum energy density $ \epsilon_0 $ is given purely by the electrostatic energy term (we ignore, or renormalise to zero, the energy of the Dirac sea). Hence
\begin{equation}
\
\epsilon_0 = \left\{  \begin{array}{ll}
\frac{1}{2} F^2 = g^2 \theta/(8 \pi^2) & (\theta \leq \pi)  \\
\frac{1}{2} (g-F)^2 = g^2 (2 \pi - \theta)^2/(8\pi^{2})  &
	 (\pi \leq \theta \leq 2 \pi).
\end{array}
\right.
\end{equation}
Thus there is a discontinuity in the slope of the energy density, corresponding
 to a first-order phase transition, at $ \theta = \pi $. In the strong-coupling
 limit $ m/g = 0 $, on the other hand, chiral invariance demands that the vacuum energy density remains {\it constant} as a function of $ \theta $ (see Sec. \ref{secboson}). Thus we expect a first-order transition at $ \theta = \pi $ for large $ m/g $, which terminates at a second order critical point at some finite $ (m/g)_c $, as illustrated in Fig. \ref{fig:phase}. This behavior was demonstrated numerically some time ago by Hamer {\it et al.} \cite{hamer82}, who located the critical point at $ (m/g)_c = 0.325(20) $, with an associated critical index $ \nu = 0.9(1) $. 

Normally, charge is confined in the model: there is a `string' of constant electric field (or flux) connecting any pair of opposite charges \cite{coleman75,casher74}. But Coleman \cite{coleman76} points out that in the very special case $ \theta = \pi $, or $ F = g/2 $, the peculiar phenomenon of ``half-asymptotic'' particles arises. In the weak-coupling limit, one can envisage the state shown in Fig. \ref{fig:halfass}. The electric field energy density is the same in between each pair of particles, and they can therefore move freely, as long as they maintain the same {\it ordering}, i.e. no pair of fermions interchanges positions.

\subsection{Boson formulation} \label{secboson}

The one-dimensional fermionic theory can be mapped into an equivalent Bose form \cite{coleman75,mandelstam75}. Some of the relevant mappings are
\begin{eqnarray}
\nonumber
: \bar{\psi} \psi : & \longleftrightarrow & -cMN_{M} \cos (2 \sqrt{\pi} \phi)  \\
\nonumber
: i \bar{\psi} \gamma_5 \psi : & \longleftrightarrow & -cMN_{M} 
\sin (2 \sqrt{\pi} \phi)  \\
j^\mu = : \bar{\psi} \gamma^\mu \psi : & \longleftrightarrow & 
\frac{1}{\sqrt{\pi}} \epsilon_{\mu \nu} \partial^\mu \phi
\label{eq12}
\end{eqnarray}
The Hamiltonian density in the charge zero sector can then be written \cite{coleman76}
\begin{equation} \label{boseham}
{\cal H} = N_M \left[ \frac{1}{2} \Pi^2 + \frac{1}{2}( \partial_1 \phi)^2 +
\frac{1}{2} M^2 \phi^2 - cm M \cos (2 \sqrt{\pi} \phi -\theta) \right]
\end{equation}
Here $ N_M $ denotes normal ordering with respect to mass $ M $, $ \phi $ is the Bose field and $ \Pi $ is its conjugate momentum, and
\begin{equation}
\nonumber
M^2 = \frac{g^2}{\pi},
\end{equation}
while%
\begin{equation}
c = \frac{e^\gamma}{2 \pi}
\end{equation}
and $ \gamma = 0.5774 $ is Euler's constant.
Note that
\begin{equation}
\partial_1 \phi = \sqrt{\pi} j^0
\end{equation}
is Gauss' law, which identifies $ \phi $ is proportional to the electric field. 
Finally, $ \theta $ is the background field variable, as before. This form of the theory is most convenient for discussing the strong-coupling limit, $ m/g \rightarrow 0 $. 

A nice discussion of the connection between $ \theta $ and chiral invariance is given by Creutz \cite{creutz95}. Consider the fermion mass term
\begin{equation}
{\cal L}_m = m \bar{\psi} \psi.
\end{equation}
If we now consider a chiral rotation
\begin{equation}
\psi \rightarrow e^{i \theta \gamma_5 /2} \psi
\end{equation}
the mass term becomes
\begin{equation}
{\cal L}_m = m' \bar{\psi} \psi + m_5 i \bar{\psi} \gamma_5 \psi
\end{equation}
where
\begin{eqnarray}
\nonumber
m' & = & m \cos \theta \\
m_5 & = & m \sin \theta,
\end{eqnarray}
while the remaining terms of the Lagrangian density are naively invariant. In 
the Bose form of the theory, the transformed mass term maps into
\begin{equation}
{\cal L}_m \rightarrow -c m M N_M \cos( 2 \sqrt{\pi} \phi - \theta )
\end{equation}
which is precisely the form (\ref{boseham}), with the chiral rotation parameter $ \theta $ playing the role of the background field.

It can be seen immediately from (\ref{boseham}) that at $ m/g = 0 $, the 
Hamiltonian is independent of the background field variable $ \theta $
(i.e. is chiral invariant), and reduces to a theory of free, massive bosons, 
with mass $ M = g/\sqrt{\pi} $ \cite{schwinger62,lowenstein71}. There is no sign of the ``half-asymptotic'' particles found in the weak-coupling limit at $ \theta = \pi $. 

Coleman \cite{coleman76} gives a very neat semi-classical argument to show how the half-asymptotic particles arise in the Bose formulation. At $ \theta = \pi $, the Hamiltonian corresponds to an effective potential
\begin{equation}
U(\phi) = \frac{1}{2} M^2 \phi^2 + c m M \cos (2 \sqrt{\pi} \phi ).
\end{equation}
For $ m/g $ small, there is a unique vacuum at $ \phi = 0 $. For $ m/g $ large,
 however, there are two vacua, located at $ \phi = \pm \frac{1}{2} \sqrt{\pi}
 $, and the symmetry $ \phi \leftrightarrow -\phi $ suffers spontaneous 
breakdown. The two vacua correspond to background field values $ \theta = 0 $ 
or $ 2 \pi $, or $ F = 0 $ or $ g $. 
Creutz\cite{creutz95} notes that in the broken symmetry phase the
expectation value of $\phi$ (and hence $\sin(2\sqrt{\pi}\phi)$) should
be non-zero, and therefore $\bar\psi\gamma^{5}\psi$ should make a
suitable order parameter, by Eq. (\ref{eq12}).

Spontaneous breakdown of a $ Z_2 $ symmetry implies that the critical point should belong to the universality class of the (1+1)D or 2D Ising model, with critical indices $ \nu = 1 $ and $ \beta = 1/8 $, which is consistent with the value for $ \nu $ found by Hamer {\it et al.} \cite{hamer82}. 

For a single scalar field in two-dimensional space-time undergoing spontaneous 
symmetry breaking, there will exist time-independent finite-energy soliton 
solutions of the classical field equations passing monotonically from one 
ground state to the other. We may designate the increasing solution a ``kink'' 
and the decreasing solution an ``antikink'' (see Fig. \ref{fig:kink}). These are the half-asymptotic 
particles, in Bose language. Kinks and antikinks must alternate with each other when well separated, just as the half-asymptotic fermions do. 

At strong couplings, the mass of the single boson excitation at $ \theta = \pi $
 can be estimated semiclassically by fitting a harmonic oscillator to the 
effective potential at $ \phi = 0 $. Hence one finds
\begin{eqnarray}
\nonumber
\frac{\Delta_2}{g} & = & \frac{M}{g} \left[ 1 - \sqrt{\pi} e^\gamma 
\left(\frac{m}{g} \right)
+ O\left(\frac{m}{g} \right)^2 \right] \\
\nonumber
& = & \frac{1}{\sqrt{\pi}} - e^\gamma\left(\frac{m}{g} \right) + 
O \left(\frac{m}{g} \right)^2 \\
\label{harmosc}
& = & 0.564 - 1.78\left(\frac{m}{g} \right) + 
O\left( \frac{m}{g} \right)^2,
\end{eqnarray}
where we have denoted this mass by $\Delta_{2}$, standing for the mass
gap in the `2-particle', or electron-positron, sector.
A crude linear extrapolation in $ m/g $ would give the boson mass vanishing at $ m/g = 0.317 $, quite close to the critical point found by Hamer {\it et al.} \cite{hamer82}.

\subsection{Lattice formulation}

We employ the Kogut-Susskind \cite{kogut75,carroll76} Hamiltonian spatial lattice formulation of the Schwinger model, with the fermions sited on a ``staggered'' spatial lattice. Let the lattice spacing be $ a $, and label the sites of the one-dimensional chain with an integer $ n $. Define a single-component fermion field $ \phi (n) $ at each site $ n $, and a link variable 
\begin{equation}
U(n,n+1) = e^{i \theta (n) } = e^{ -i a g A^1 (n)}
\end{equation}
on each link\footnote{We trust the index $ (n) $ will differentiate the lattice fields $ \phi(n) $ and $ \theta(n) $ from the quite different fields $ \phi $ and $ \theta $ of Sec. \ref{secboson}}. Then the lattice Hamiltonian equivalent to (\ref{fermionham}) is
\begin{equation} \label{latticeham}
H = - \frac{i}{2a} \sum_{n=1}^{N} \left[ \phi^\dagger(n) e^{i \theta(n)}
\phi (n+1) -\mbox{h.c.} \right]
+ m \sum_{n=1}^{N} (-1)^n \phi^\dagger (n) \phi (n)
+ \frac{g^2 a}{2} \sum_{n=1}^{N} L^2 (n)
\end{equation}
where the number of lattice sites $ N $ is even, and the correspondence between lattice and continuum fields is
\begin{equation}
\phi(n)/\sqrt{a} \rightarrow \left\{ 
\begin{array}{ll}
\psi_{\mbox{\tiny upper}} (x)  & \ \ \ \   n \mbox{ even} \\
\psi_{\mbox{\tiny lower}} (x)  & \ \ \ \   n \mbox{ odd}
\end{array}
\right.
\end{equation}
\begin{eqnarray}
\frac{1}{ag} \theta(n) & \rightarrow & - A^1 (x) \\
g L(n) & \rightarrow & E(x) .
\end{eqnarray}
The $ \gamma $ matrices are represented by
\begin{equation}
\gamma^0 = \left( 
\begin{array}{cc}
1 & 0 \\
0 & -1
\end{array}
\right),
\ \ \ \ \ 
\gamma^1 = \left( 
\begin{array}{cc}
0 & 1 \\
-1 & 0
\end{array}
\right).
\end{equation}
We use a ``compact'' formulation where the gauge field becomes an angular variable $ 0 \leq \theta(n) \leq 2 \pi $ on the lattice, and $ L(n) $ is the conjugate spin variable
\begin{equation}
[ \theta(n), L(m) ] = i \delta_{nm}
\end{equation}
so that $ L(n) $ has integer eigenvalues $ L(n) = 0, \pm 1, \pm 2, \dots $. In the naive continuum limit $ a \rightarrow 0 $, the lattice Hamiltonian (\ref{latticeham}) reduces to the continuum expression (\ref{fermionham}).

The Hamiltonian is transcribed to a dimensionless operator
\begin{equation}
W = \frac{2}{a g^2} H = W_0 + x V
\end{equation}
where
\begin{eqnarray}
W_0 & = & \sum_n L^2 (n) + \mu \sum_n (-1)^n \phi^\dagger (n) \phi (n), \\
V & = & -i \sum_n \left[ \phi^\dagger (n) e^{i \theta(n) } \phi(n+1)
- \mbox{h.c.} \right]
\end{eqnarray}
\begin{equation}
\mu = \frac{2m}{g^2 a}, \ \ \ \ \ x = \frac{1}{g^2 a^2}.
\end{equation}

In the lattice strong-coupling limit $ x \rightarrow 0 $, the unperturbed ground state $ | 0 \rangle $ has
\begin{equation}
\label{groundalpha0}
L(n) = 0, \ \ \ \ \ 
\phi^\dagger(n) \phi(n) = \frac{1}{2} [ 1 - (-1)^n ], \ \ \ \ \ 
\mbox{all } n
\end{equation}
whose energy we normalise to zero. The lattice version of Gauss' law is then taken as
\begin{equation} \label{latticegauss}
L(n) - L(n-1) = \phi^\dagger (n) \phi(n) - \frac{1}{2}[1-(-1)^n]
\end{equation}
which means excitations on odd and even sites create $ \mp 1 $ units of flux, corresponding to ``electron'' and ``positron'' excitations respectively. Eq. (\ref{latticegauss}) determines the electric field $ L(n) $ entirely, up to an arbitrary additive {\it constant} $ \alpha $, which then represents the background field. Allowing $ \alpha $ to be non-zero, the electrostatic energy term is modified to
\begin{equation}
\sum_n L^2 (n) \rightarrow \sum_n (L(n) + \alpha)^2 .
\end{equation}
The physics of the background field then matches precisely with the continuum 
discussion. If $ | \alpha | > 1/2 $, charged pairs will be produced and separate to infinity to lower the electrostatic energy, and bring 
$ | \alpha | \leq 1/2 $. Physics is then periodic in $ \alpha $ with period 1, and the background field variable is
\begin{equation}
\theta = 2 \pi \alpha .
\end{equation}

In the weak-coupling limit $ m/g \rightarrow \infty $ the vacuum contains no fermion excitations, and the ground state energy at $ x=0 $ is
\begin{equation}
\omega_0 = N \alpha^2 \ \ \ \ \ \ \ \ \  
( |\alpha | < 1/2 )
\label{eq39}
\end{equation}
corresponding to the `zero-loop' state with $ \{ L(n) = 0, $ all $ n  \} $; or
\begin{equation}
\omega_0 = N (1-\alpha)^2 \ \ \ \ \ 
(1/2 < \alpha < 1 )
\label{eq40}
\end{equation}
corresponding to the `one-loop' state with $ \{ L(n) = -1,  \mbox{ all } n  \} $. 

At $\alpha = 1/2 $, these two states are degenerate; and half-asymptotic states
 consisting of alternating electrons and positrons exist, just as in the 
continuum model. Let us denote the two states (\ref{eq39}) and (\ref{eq40}) by 
$ | \frac{1}{2} \rangle $ and $ | -\frac{1}{2} \rangle $, according to the 
electric fields $ \{ L(n) + \alpha = \pm \frac{1}{2}, \mbox{ all } n \} $ on 
the links. On a finite lattice at weak coupling, the eigenstates will be
the symmetric and antisymmetric combinations
\begin{equation}
| 0 \rangle = \frac{1}{\sqrt{2}} [ | \mbox{$\frac{1}{2}$} \rangle + | -\mbox{$\frac{1}{2}$} \rangle ]
\end{equation}
and
\begin{equation}
\label{loopstate}
| 0'  \rangle = \frac{1}{\sqrt{2}} [ | \mbox{$\frac{1}{2}$} \rangle -  | 
-\mbox{$\frac{1}{2}$} \rangle ].
\end{equation}
These states become degenerate in the bulk limit $N \rightarrow \infty$,
when spontaneous symmetry-breaking occurs. We shall denote the energy
gap between them (or ``loop gap") as $\Delta_{0}$, standing for the
0-particle sector.

The lowest-energy single-particle excitation (Fig. \ref{fig:halfasslone}) is
\begin{equation}
|1 \rangle = \sqrt{\frac{2}{N}} 
 \sum_{ \scriptsize
\begin{array}{cc}
n = 1 \\
\{n \mbox{ odd} \}
\end{array}
}^{N}
\phi (n) \prod_{j=n}^{N} e^{-i \theta (j)} |\mbox{$\frac{1}{2}$} \rangle
\end{equation}
while the lowest ``2-particle'' state is 
\begin{equation}
|2 \rangle =  \frac{1}{\sqrt{N}} 
\sum_{ \scriptsize
\begin{array}{cc}
n = 1 \\
\{n \mbox{ odd} \}
\end{array}
}^{N} 
\left[
\phi^{\dagger}(n+1) e^{- i \theta(n)} \phi(n) | \mbox{$\frac{1}{2}$} \rangle + 
\phi^{\dagger}(n+1) e^{i \theta(n)} \phi(n+2) |-\mbox{$\frac{1}{2}$} \rangle
\right].
\end{equation}
We shall compute the energy gaps between these states and the ground state,
$\Delta_{1}/g$ and $\Delta_{2}/g$ respectively.

We will also study two order parameters which can be used to
characterize the phase transition at $\theta = \pi$. The first one is
the average electric field
\begin{equation}
\Gamma^{\alpha} = \frac{1}{N} \langle \sum_{n}(L(n) + \alpha) \rangle_{0},
\end{equation}
which in the weak-coupling limit $m/g \rightarrow \infty$ takes values
$\pm 1/2$ for the zero-loop and one-loop states respectively. By the
Feynman-Hellman theorem, it is proportional to the slope
$\partial E_{0}/\partial \alpha$ of the ground-state energy: a glance at
Fig. \ref{fig:phase} shows that at large $m/g$ this slope 
undergoes a discontinuity at $ \theta = \pi$, while below $(m/g)_{c}$ it vanishes.

	The second order parameter is the axial fermion density
suggested by Creutz\cite{creutz95}
\begin{eqnarray}
\Gamma^{5} &  = &  \langle i \bar\psi\gamma^{5}\psi/g \rangle_{0} \\
 & = & - \frac{i \sqrt{x}}{N} \langle \sum_{n} (-1)^{n} \left[ \phi^{\dagger}(n)\phi(n+1)
- \mbox{h.c.} \right] \rangle_{0}
\end{eqnarray}
where
\begin{equation}
\gamma^{5} = i\gamma^{0}\gamma^{1}.
\end{equation}

	Now on a finite lattice there is no spontaneous symmetry
breakdown, and the expectation values of the order parameters will
remain identically zero. The remedy for this is well-known,
however\cite{yang}. If the two lowest-energy states which become
degenerate in the bulk limit $ N \rightarrow \infty$ are denoted $|0 \rangle$
and $|0' \rangle $ respectively, then a simple $ 2 \times 2 $ matrix calculation shows that
the order parameter corresponding to an operator $Q$ can be estimated as
the {\it overlap} matrix element $ \langle 0|Q|0' \rangle $ on the finite lattice.

In zero background field $\alpha =0$, the 2-particle `positronium' excited 
states of lowest energy in the lattice strong coupling limit are the 
`vector' state \cite{banks76}
\begin{equation}
| v \rangle  = \frac{1}{\sqrt{N}} \sum_{n=1}^{N}
\left[
\phi^{\dagger}(n) e^{i \theta(n)} \phi(n+1) + \mbox{h.c.}
\right]
|0 \rangle 
\end{equation}
and the `scalar' state
\begin{equation}
| s \rangle  = \frac{1}{\sqrt{N}}\sum_{n=1}^{N}
\left[
\phi^{\dagger}(n)e^{i \theta(n)} \phi(n+1) - \mbox{h.c.}
\right]
|0 \rangle
\end{equation}
where $ |0 \rangle $ is the strong-coupling ground state (\ref{groundalpha0}).

\subsection{Lattice spin formulation}

An equivalent lattice spin formulation can be obtained by a Jordan-Wigner transformation \cite{banks76}
\begin{eqnarray}
\phi(n) & = & \prod_{l<n} [ i \sigma_3 (l) ] \sigma^-(n) \\
\phi^\dagger (n) & = & \prod_{l<n} [ -i \sigma_3 (l) ] \sigma^+(n) 
\end{eqnarray}
giving
\begin{eqnarray}
\label{spinham1}
W_0 & = & \sum_n ( L(n) + \alpha)^2 + 
\frac{\mu}{2} \sum_n  (-1)^n \sigma_3(n) + \frac{N \mu}{2} \\
\label{spinham2}
V & = & \sum_n \left[ \sigma^+ (n) e^{i \theta (n)} \sigma^-(n+1) 
+ \mbox{h.c.} \right]
\end{eqnarray}
and
\begin{equation}
\label{spingauss}
L(n) - L(n-1) = \frac{1}{2}[ \sigma_3(n) + (-1)^n ].
\end{equation}
The $ \Gamma^5 $ order parameter can be written in spin variables
\begin{equation}
\Gamma^5 = \frac{\sqrt{x}}{N} \langle \sum_n (-1)^n 
\left[ \sigma^+ (n) e^{i \theta (n)} \sigma^-(n+1) 
+ \mbox{h.c.} \right] \rangle_0 .
\end{equation} 
This is the form which we used in the numerical calculations.

\section{METHODS}

\subsection{The density matrix renormalisation group method}

Our results are based on the density matrix renormalisation group (DMRG) method \cite{white92,gehring97}. DMRG has been used primarily to study low dimensional
 quantum lattice systems in condensed matter physics, and is able to obtain 
with great accuracy quantities such as the ground and excited state energies, 
and correlation functions. One of the key features of DMRG is its ability to 
calculate these quantities for very large system sizes, particularly for 
systems with a low number of degrees of freedom per site. 

The method employed here is the ``infinite system'' DMRG method, as prescribed 
by White \cite{white92}, used both with open and periodic boundary conditions
(OBC and PBC).  Due to the presence of the electric field on the links, and the
 differing nature of the odd and even-numbered sites, some modifications were 
made to the method, although nothing that changes the spirit of the DMRG. The 
quantities calculated are the ground state energies, mass gaps and order 
parameters. We use the form of the Hamiltonian given in (\ref{spinham1}) and 
(\ref{spinham2}). 

Let us first look at how the presence of the electric fields affects the 
implementation of the code. For a particular spin configuration with OBC, 
if we specify the incoming electric field for the first site of the chain, 
then according to (\ref{spingauss}) the electric field for all the links can 
be deduced. This incoming field can be either $ L_{\tiny \mbox{in}} = 0 $ for 
the zero background field case, or $ L_{\tiny \mbox{in}} = \pm 1/2 $ for the 
background field case. If PBC are imposed, then as there is no particular link 
to fix the electric field, we can have loops of electric flux extending 
throughout the ring. In the presence of a background field we simply add (or 
subtract) 1/2 from the values of the electric fields due to the spin 
configuration. A cutoff, or maximum loop value was chosen such that full convergence was 
reached to machine precision. A loop cutoff of [-5,5] was more than sufficient 
in most cases.

A typical DMRG iteration is shown in Figs. \ref{fig:dmrgobc} (OBC) and 
\ref{fig:dmrgpbc} (PBC). The chain (or ring) is split into two blocks and two sites, where blocks contain in general more than one site. In one DMRG iteration, we augment two sites to each block, so that the whole system grows by 4 sites each time. This is more convenient than the standard approach of adding one site at a time, due to the differing nature of the odd and even-numbered sites. For the PBC case this allows one to have the same type of site on each end of the block (i.e. both even or both odd), which considerably simplifies the bookkeeping of the electric fields on the links. In this case both blocks are identical, so that only one augmentation and density matrix is necessary. For the OBC case, we have a different situation. Again augmenting two sites at a time allows us to have the same type of site on each end of the blocks, but in this case the left hand block will always have an odd site on each end, while the right hand block will have an even site on each end. This requires two separate density matrices for each block, as they are not identical. 

We start with a 4 site superblock, and continue adding sites until we reach 
256 sites. For PBC we keep a maximum of 125 states per spin sector per loop 
value in a block, which corresponds to a truncation parameter of approximately 
$ l \sim 1100 $, where $l$ is the total number of basis states
retained in a block. The truncation error for 256 sites can vary from 1 part in
$10^{7}$ to $10^{13}$ depending on the number of excitations and the 
parameters. 
For OBC, again we keep a maximum of 125 states per spin sector, but as there 
are no electric field loops, here the truncation parameter is $ l \sim 300 $. 
The truncation error for 256 sites is typically 1 part in $10^{16}$. 

\subsection{Accuracy of the DMRG calculations}

The accuracy of the DMRG calculation is strongly dependent on the parameter 
values $ x $ and $ m/g $; in particular we expect the accuracy to worsen near 
critical regions. We find that OBC provide more accurate results than PBC, which is 
consistent with previous DMRG studies. We provide details of the accuracy for 
the worst case scenario, i.e. the parameters closest to the critical regions. 
This is not representative of the typical accuracies achievable with the DMRG, 
but gives us an upper bound on the errors. 

We examine the point $ (m/g = 0.3; x = 100; \theta = \pi; N = 256) $ with PBC, 
which represents the smallest lattice spacing used for this study, and lies in 
the critical region. Table  \ref{tab:convpbcb0.5} shows the convergence of the 
DMRG for these parameters as a function of the truncation parameter $ l $,  
for the energies and order parameters. The ground state energy can be resolved 
to 1 part in $10^{6}$, while the two-particle gap is resolved to 1 part
in $10^{3}$. The order parameters are subject to round-off errors because
they involve the overlap of the two wavefunctions, and hence reduced accuracy 
is expected: here we have 3 figures. Fig. \ref{fig:dmrgconvpbc} shows the 
behaviour of the ground state energy density as a function of the density 
matrix truncation eigenvalue, which corresponds to the sum of the eigenvalues 
thrown away from the density matrix. We see a linear relationship between these
 quantities, which is confirmation that the DMRG is working correctly. Where 
possible, we obtain the final estimate by taking the results with the two 
largest truncation parameters, and performing a linear extrapolation to the 
$y$-axis. The error estimate is obtained by taking the difference in the 
extrapolant and the result with the largest $ l $. 

Using OBC improves the accuracy of the DMRG calculation considerably. Table 
\ref{tab:convobcb0.5} shows the convergence of the same quantities with the 
same parameters as Table 
  \ref{tab:convpbcb0.5}  
, but with OBC. The ground state energy density is accurate to nearly machine  
precision, and the 2-particle mass gap to 1 part in $10^{7}$. Again due to 
round-off errors, we are limited in accuracy for the order parameters, so a
similar accuracy to the PBC code is achieved here. A point against using
OBC is that the finite-size corrections are much larger for this case. We use 
OBC to calculate the 1-particle gap and the order parameter estimates in the 
continuum limit, and PBC otherwise.

\section{RESULTS AT BACKGROUND FIELD $ \theta = \pi $}

\subsection{Analysis of critical behaviour}

We use finite-size scaling theory to estimate the position of the critical point, by calculating pseudo-critical points \cite{barber83} at each lattice size $ N $, and each lattice spacing $ x $. We demand that the scaled energy-gap ratio
\begin{equation}
R_N (m/g) = \frac{N \Delta_N (m/g)}{(N-1) \Delta_{N-1} (m/g)}
\end{equation}
is equal to unity at the point $ m/g = (m/g)_N^* $, which is the pseudo-critical
 point. Here $ \Delta_N (m/g) $ refers to the energy gap at some finite lattice size
 $ N $.
In practice we calculate $ R_N 
(m/g) $ for a cluster of five points straddling the pseudo-critical point, 
then use a polynomial interpolation to find $ (m/g)_N^* $. The points were 
chosen at a spacing of $ \Delta(m/g) = 0.02 $ apart, the smallest reasonable 
spacing which ensures we cover the pseudo-critical point based on the work of 
Hamer {\it et al.} \cite{hamer82}. For this exercise we use the `loop
gap' $\Delta_{0}/g$, which collapses to zero at the pseudo-critical point. 

Fig. \ref{fig:pseudoplot} shows three sample data sets for the pseudo-critical points calculated between $ x = 4 $ and $ x = 100 $. On a $ 1/N^3 $ plot these 
can be simply extrapolated with a quadratic, with errors in the vicinity 
of 1 part in $10^{4}$, for all $ x $. Some numerical instability creeps in for 
the larger lattice sizes, however, where the change in the gap energy with 
lattice size $ N $ becomes so slow that round off errors become appreciable. 
The continuum limit is then estimated from these bulk critical points, as 
 shown in Fig. \ref{fig:finalplot}. 
A quadratic fit in $ 1/\sqrt{x} $ extracts 
the continuum limit $ a \rightarrow 0$ or $ x \rightarrow \infty$, which we estimate to be
\begin{equation}
\label{criticalpoint}
\left( \frac{m}{g} \right)_c = 0.3335(2).
\end{equation}
This is consistent with the previous estimate by Hamer {\it et al.}\cite{hamer82}
 of $ (m/g)_c
 = 0.325(20) $, or Schiller and Ranft\cite{schiller82}, $(m/g)_{c} =
0.31(1)$, but with two orders of magnitude improvement in accuracy. 

Finite-size scaling theory  \cite{barber83} also allows us to estimate the 
critical indices for the model. The theory tells us that the Callan-Symanzik 
``beta function''
\begin{equation}
\frac{ \beta_N(m/g)}{m/g} =  \frac{ \Delta_N ( m/g)}{( \Delta_N (m/g) - 2 (m/g) \Delta_N ' (m/g))} 
\end{equation}
scales at the critical point like $ \beta_N ( (m/g)_c ) \sim N^{-1/\nu} $ as $ N \rightarrow \infty $ \cite{hamer81}. Hence we can extract the critical exponent $ \nu $ from the ratio 
\begin{equation}
\label{ratiolinear}
N \left( 1 - \frac{\beta_N ( (m/g)_N^* ) }{ \beta_{N-1} ( (m/g)_{N-1}^* )} \right), \end{equation}
since this should approach $ 1/\nu $ as $ N \rightarrow \infty $. Alternatively we can use the limiting behavior
\begin{equation}
\label{ratiolog}
\frac{ \ln [ \beta_N ( (m/g)_N^* ) / \beta_{N-1} ( (m/g)_N^* ) ] }
{ \ln [ N/(N-1) ] } \sim - \frac{1}{\nu} 
\end{equation}
as $ N \rightarrow \infty $. We call the first of these ratios 
(\ref{ratiolinear}) the `linear' estimate, while the second is the `logarithmic'
 estimate. We calculate these ratios for all lattice sizes $ N $ and lattice 
spacings $ x $. Analysis of the data shows that the `logarithmic' version is 
more stable numerically, and converges more quickly towards the bulk limit. We 
show an example of this convergence in Fig. \ref{fig:ratios} for a particular 
lattice spacing $ 1/\sqrt{x} = ga = 0.45 $. We see monotonic convergence towards
 $ - 1/\nu \rightarrow -1 $, until the same numerical errors begin to
creep in as seen in Fig. \ref{fig:pseudoplot}. 

A similar method was used to estimate the critical index $ \beta $. The
order parameters $ \Gamma^\alpha_N $ and $ \Gamma^5_N $ are expected to scale like 
$ \sim N^{-\beta/\nu} $ \cite{hamer81}. `Linear' and `logarithmic' ratios are 
again constructed from these quantities, and analysed in the same way as for 
the ``beta-functions''. Again the logarithmic ratios 
\begin{equation}
\frac{ \ln [ \Gamma_N ( (m/g)_N^* ) / \Gamma_{N-1} ( (m/g)_N^* ) ] }
{ \ln [ N/(N-1) ] } \sim - \frac{\beta}{\nu} 
\end{equation}
seem to do better than the `linear' ratios, in terms of numerical stability 
and convergence. Fig. \ref{fig:ratios} shows the estimate for $ \beta/\nu $, 
which uses the electric field order parameter. All final estimates for 
$ \beta/\nu $ shown in Table \ref{tab:critexp} are calculated using 
$ \Gamma_N^\alpha = \langle (L + \alpha ) \rangle_0 $. 

Table \ref{tab:critexp} displays
 our results for the critical exponents, for each different lattice spacing. We
 see essentially no variation in the exponents with lattice spacing, to within
the accuracy of our calculations. 
Our best estimates for the 
critical exponents are thus
\begin{eqnarray}
\nu & = & 0.99(1) \\
\beta/\nu & = & 0.125(5) .
\end{eqnarray}
These results provide reasonably conclusive evidence that the Schwinger model
transition at 
$ \theta = \pi$ lies in the same universality class as the one dimensional 
transverse Ising model, or equivalently the 2D Ising model, with ($ \nu = 1, 
\beta = 1/8 $).

\subsection{Behaviour in the continuum limit}

We now turn to estimating continuum limit values for the energy gaps and 
order parameters.  
First of all, examination of the data points for 2-particle gaps in Fig. 
\ref{fig:bulk2part} 
reveal that some extrapolation is necessary, especially for small lattice 
spacings, to obtain the bulk limit $N \rightarrow \infty$. 
In order to do so, we need to fit the data with an appropriate function
of $ N $. 
Now results from the last section revealed that the critical behavior of this 
model is closely related to that of the transverse Ising model. In that
model the finite lattice values away from the critical point converge
exponentially to the bulk limit, modulo half-integral powers of
$N$\cite{hamer81} (see appendix).
%
%
%
Hence we apply a form
\begin{equation}
\label{fitfunc}
\Delta_{N} = \left( a + \frac{b}{\sqrt{N}} + \frac{c}{N} \right) \exp (-d N) + e
\label{eq55}
\end{equation}
to the data, where $ a,b,c,d,e $ are fit parameters. Fig. \ref{fig:bulk2part} 
shows that this form in fact fits the data very well, for all values of $ m/g $. 

As a double-check of the results of the fit, we also apply a VBS sequence 
extrapolation routine \cite{vandenbroeck79} to the finite-lattice
sequences, which provides an independent estimate of the bulk limit. Errors 
for the VBS estimates were obtained by examining the columns of the VBS 
extrapolants, and comparing the results with
different VBS parameters $ \alpha $. Typically the VBS and the fit
results agree to better than 1\% accuracy, as long as we are not near the 
critical region where it becomes difficult to extract a reliable estimate. 

Fig. \ref{fig:cont2part} shows the extrapolation to the continuum limit $x
\rightarrow \infty$, or $ a \rightarrow 0$.
We see that generally there is good agreement 
between the VBS extrapolations and the fit (\ref{eq55}), except towards small 
lattice spacings where a discrepancy opens up. 
An extrapolation to the continuum limit is now performed by a simple
polynomial fit in powers of 
$1 / \sqrt{x} =ga $.  
The double extrapolation therefore introduces quite large errors into our final
 results, compared to the original DMRG eigenvalues where the errors are at 
worst of order $ 0.1 \% $.

\subsubsection{Loop energies}

Fig. \ref{fig:mogland} shows our results for the ``loop'' energy gap
$\Delta_{0}/g$ for all values of $ m/g $. We see that this gap vanishes at the 
same point as calculated in (\ref{criticalpoint}), and is zero for all $ m/g > 
(m/g)_c $, as predicted by Coleman \cite{coleman76}. The convergence to the 
bulk limit $ N \rightarrow \infty $ for this case is much better than for the 
2-particle gap illustrated in Fig. \ref{fig:bulk2part}, such that away from 
the critical region hardly any extrapolation is needed. This gives us good 
accuracy for the region $ m/g < 0.25 $, where the gaps are resolved to 4 
figures. Near the critical region it becomes rather more difficult to 
extrapolate in a consistent manner.

\subsubsection{2-particle gap}

We plot on the same figure results for the 2-particle gap $\Delta_{2}/g$. 
Again, the gap vanishes at the critical point $ (m/g)_c $, but on either side of the critical point there is a finite gap, and an
almost linear behaviour with $ m/g $. A linear fit through the points in the 
range $ m/g = [0,(m/g)_c] $ gives 
\begin{equation}
\frac{\Delta_2}{g} \approx 0.569 - 1.72 \left(\frac{m}{g} \right),
\end{equation}
which agrees very well with the prediction made in (\ref{harmosc}). Note,
however, 
that the behaviour is not {\it exactly} linear. 

	The $ m/g = 0 $ case is unique in the sense that we have a direct check
 on our results, since an analytic solution is known. For the massless case  
the background field has no effect on the physics of the model, since
any background field is completely screened out.
Hence all measurable quantities should be independent of the background
field, and in particular the Schwinger boson mass \cite{schwinger62,lowenstein71}
\begin{equation}
\label{exactresult}
\frac{\Delta_2}{g} = \frac{1}{\sqrt{\pi}} \approx 0.56419.
\end{equation}
Our DMRG estimate gives $ \Delta_2 /g = 0.57(1) $, which agrees with this result
within errors. 

\subsubsection{1-particle gap}

The 1-particle gap must be calculated using open boundary conditions, due to 
the mismatch in electric fields at either end of the chain. Fig. 
\ref{fig:halfasslone} shows the situation. A single charged particle in
the middle will shift the electric field by $ \pm g $, and hence we have 
differing electric fields at either end. There is a further complication in 
the case with open boundaries, in that applying a background field of either 
$ \alpha = 1/2 $ or $ -1/2 $ gives a different result for the {\it ground
state energy } 
for a finite lattice. This is due to the `staggered
lattice' convention we have adopted, with `electrons' appearing on odd
sites, and `positrons' appearing on even sites.
The only consistent definition of the one-particle gap for a finite lattice 
is therefore
\begin{equation}
\Delta_1/g = E_1 - \frac{1}{2}( E_0^+ + E_0^-)
\end{equation}
where $ E_1 $ is the 1-particle energy, $ E_0^+ $ is the ground state energy 
with $ \alpha = 1/2 $, and $ E_0^- $ that for $ \alpha = -1/2 $.

Apart from this complication the procedure in extrapolating to the continuum 
limit is the same as for the 2-particle state. Our results are shown in 
Fig. \ref{fig:mogland}. We see that the gap vanishes for $ m/g < (m/g)_c $,
while 
for $ m/g > (m/g)_c $ the 1-particle gap is very close to half the 2-particle 
gap. 
Once again, the behaviour is very nearly linear in $m/g$.

	The pattern of eigenvalues exhibited in Fig. \ref{fig:mogland} bears
an extraordinary resemblance to that of the transverse Ising
model\cite{fradkin,hamer81}, even down to the (almost) linear behaviour with
$m/g$. In particular, we see that the energy of the 1-particle or `kink'
state vanishes at the critical point, and then {\it remains degenerate}
with the ground state for $(m/g) < (m/g)_{c}$. Assuming this degeneracy
is exact, this indicates that a `kink condensate' will form in the
ground state for small mass, as discussed by Fradkin and
Susskind\cite{fradkin}. It also indicates the existence of a `dual
symmetry' in the model, additional to that discussed in Section II,
which is spontaneously broken in the low-mass region, and has not been 
explored hitherto. There should also be a `dual order
parameter' associated with this symmetry.
By analogy with the Ising case, one would expect the 'dual' order
parameter to be the expectation value of the kink creation/destruction
operator, and the 'dual' symmetry to correspond to inversion of this
operator; but we have not verified this by explicit computation.

	The other notable feature of Fig. \ref{fig:mogland} is the
degeneracy (within errors) between the zero-particle gap $\Delta_{0}/g$
and the 2-particle gap $\Delta_{2}/g$ at small mass. Our physical
pictures based on a weak-coupling representation do not give clear
physical insight into this phenomenon.

	All the features just discussed are of course peculiar to the
special case $\theta =\pi$. For instance, the 1-particle gap at any other 
value of $\theta$ would be infinite.
 
\subsubsection{Order Parameters}

We may also obtain estimates for the order parameters $\Gamma^5$ and
$\Gamma^\alpha$ as functions of $m/g$. Our results are displayed in Fig. 
\ref{fig:moglandorder}. Both order parameters are zero, within errors, for 
$ m/g < (m/g)_c $. Near the critical region, particularly for $ m/g < (m/g)_c $
 it becomes quite difficult to obtain accurate estimates, 
but it appears that both order parameters turn over and drop abruptly to
zero as the critical point is approached from above, consistent with the
small exponent $ \beta = 1/8 $ found in the previous section.
The axial density $\Gamma^5$ decreases steadily towards zero at large
$m/g$, whereas $\Gamma^\alpha$ approaches the expected value of $1/2$.
All our results for the gaps $ \Delta_0/g $, $ \Delta_1/g $, $ \Delta_2/g $ and both order parameters are shown in Table \ref{tab:bfieldresults} for future reference.

\section{RESULTS AT BACKGROUND FIELD $ \theta = 0 $}
\label{sec:zeroalpha}

The case of zero background field has been studied by many authors 
already, as outlined in the introduction. Our purpose here is to demonstrate
how the DMRG can improve on the accuracy of existing results. The main 
quantities of interest are the ``vector'' and ``scalar'' state 
masses. The most accurate results to date are those of Sriganesh {\it et al.}
\cite{sriganesh00}, who used numerical exact diagonalisation results together 
with a VBS extrapolation to obtain their final estimates. The largest lattice 
size calculated by these authors was $ N = 22 $, hence we expect to be able to 
do much better using our DMRG algorithm, which can go to much larger lattice 
sizes. 

For the $ \theta = 0 $ case, we find that convergence with lattice size is much
more rapid than for $ \theta = \pi  $, so much so that with $ N = 256 $, there 
is essentially no extrapolation necessary to obtain the bulk limit. 
Fig. \ref{fig:vectorm0} shows the data for $ m/g = 0 $, which is the exactly 
soluble case. For a particular lattice spacing $ x $, we have very good 
convergence with lattice size, so that for $ x = 4 $ through to $ x = 100 $ we 
have 6 figure convergence. Even for the largest lattice spacing $ x = 400 $ 
used here, the gaps have converged to 4 figures. Our final continuum estimate 
is obtained by the method of linear, quadratic and cubic extrapolants in
$1/\sqrt{x}$ as 
used in Ref. \cite{sriganesh00}. An example is shown in 
Fig. \ref{fig:linquadcubm0}. Our final estimate of $ m^-/g = 0.56419(4) $
for this case agrees extremely well with the analytic result 
(\ref{exactresult}). 

Table \ref{tab:vector} summarises our results,
which show  
between one and two orders of magnitude improvement in accuracy over the 
previous 
best estimates for small values of $ m/g $. 
We must note, however, that for values of $ m/g > 1 $ we have little or no 
improvement in accuracy over previous results. To explain this, first we must 
note that the structure of the eigenvalue function in $ x $ shifts towards 
large $ x $ for large $ m/g $. But at large $x$ and large $ m/g $ there are 
many ``intruder'' states below the vector state for finite $ N $, artifacts
of the finite lattice corresponding to states with no fermion excitations but 
with loops of electric flux winding around the entire lattice. This restricts 
the range of $ x $ that can be used, and hence the accuracy of the calculation.
 In a future calculation, a way of eliminating these intruder states must be 
found. 

A comparison of our results with previous works is shown in 
Fig. \ref{fig:moglandvector}. In the low mass region, we see excellent 
agreement between our results and the series expansion around $ m/g = 0 $. Our
results are also fairly consistent with the results of Sriganesh {\it et al.}, 
obtained by exact diagonalisation. The 'fast-moving frame' results of 
Kr\"{o}ger and Scheu seem to be consistently a little low in this region. In 
the large mass region the situation is reversed, as Kr\"{o}ger and Scheu obtain
excellent agreement with the non-relativistic expansion, while both our results
 and those of Sriganesh {\it et al.} seem to be slightly high. We attribute 
his to the problem with intruder states discussed above. The Kr\"{o}ger and 
Scheu  
quasi-light-cone approach appears to remain more accurate in this
weak-coupling region.

Scalar state mass gaps $ m^+/g $ may also be calculated using DMRG; here we merely demonstrate its applicability for one mass value $ m/g = 1.0 $, where we obtain $ m^+/g - 2m/g = 1.118(1) $. This is again over an order of magnitude better that the previous best estimate of Sriganesh {\it et al.} who obtained for the same quantity 1.12(3).

\section{CONCLUSIONS}

	In this paper, we have demonstrated the application of the
numerical DMRG technique of White and collaborators\cite{white92} to a
lattice gauge model, namely the massive Schwinger model. The results are
in most cases nearly two orders of magnitude more accurate than previous
calculations. Where exact diagonalisation can treat lattices up to 22
sites, DMRG gives very accurate results up to 256 sites. The long-range
Coulomb interaction present in the model has proved no impediment to the
DMRG technique: this is very likely connected with the fact that that
the Coulomb interaction is screened\cite{schwinger62,casher74}, and the {\it effective}
interactions are short-range.

	The most interesting results were obtained for the case of
background field $\theta =\pi$. We have performed a detailed study
confirming the existence of the `kinks' or `half-asymptotic' particles
predicted by Coleman\cite{coleman76}, and have shown that there is a phase
transition at $(m/g)_{c} = 0.3335(2)$ belonging in the universality
class of the transverse or (1+1)D Ising model. The pattern of energy
eigenvalues near the critical point bears a truly remarkable resemblance
to the transverse Ising model, and points to some new physical effects,
beyond those discussed by Coleman\cite{coleman76}.

	In particular, for $(m/g) < (m/g)_{c}$ and $\theta = \pi$, we
find that the single-kink state is {\it degenerate} with the ground
state. This points to the existence of a ``kink condensate"
\cite{fradkin} in the ground state. It also points to the existence of a
`dual symmetry' and `dual order parameter' in the model, which have not
been discussed hitherto. We also find that the gap between the symmetric
and antisymmetric `loop state' combinations appears to be exactly
degenerate with the 2-particle vector gap in this low-mass region. 

Some calculations were also carried out at zero background field,
$\theta = 0$. The vector gap was obtained with great precision for small
$m/g$; but at large $m/g$ the accuracy was spoiled a little by
finite-lattice ``intruder'' states.

	The same methods should be applicable to other lattice gauge
models in (1+1) dimensions. In higher dimensions, however, the DMRG
technique does not have such a large comparative advantage over other
techniques. Strenuous efforts are being made to develop improved DMRG
algorithms for lattice models in (2+1) D; but nothing has even been
attempted in (3+1)D, as far as we are aware.

\section{ACKNOWLEDGMENTS}

We would like to thank Professors Michael Creutz and Jaan Oitmaa for
useful discussions on this topic, and Dr. Zheng Weihong for numerical
checks of some of our results by series methods.
We are grateful for computational facilities provided by the New South Wales 
Center
for Parallel Computing, the Australian Center for Advanced Computing
and Communications and the Australian Partnership for Advanced Computing.
R.B. was supported by the Australian Research Council and the J. G. Russell Foundation.

\section{APPENDIX}

Here we calculate analytically the form of the finite-size scaling
corrections in the 1D transverse Ising model. We follow the discussion
by Hamer and Barber\cite{hamer81}, hereafter referred to as I.

The Hamiltonian is taken as

\begin{equation}
H = \sum_{m=1}^{M} (1-\sigma_{3}(m)) - x \sum_{m=1}^{M}
\sigma_{1}(m)\sigma_{1}(m+1)
\label{A1}
\end{equation}

The model can be solved exactly by the methods of Schultz, Mattis and
Lieb\cite{schultz66}.
The mass gap on a finite lattice of $ M $ sites with periodic boundary
conditions is

\begin{equation}
F(x,M) = 2(1-x) + 2M[T_{2M}(x) - T_{M}(x)]
\label{A2}
\end{equation}

 where 

\begin{equation}
T_{M}(x) = \frac{1}{M}\sum_{k=0}^{M-1} \Lambda(\frac{\pi k}{M})
\label{A3}
\end{equation}

and

\begin{equation}
\Lambda(\theta) = [(1-x)^{2} + 4x\sin^{2}\theta]^{1/2}.
\label{A4}
\end{equation}

For fixed $ x $, as $M \rightarrow \infty$,

\begin{equation}
\tilde{T}_{M}(x) = \frac{1}{\sqrt{x}}T_{M}(x) \sim c_{0} +
(\frac{4\pi}{x})(1-x^{2})^{1/2} \frac{e^{-2Mw_{c}(x)}}{(2M)^{3/2}}
\label{A5}
\end{equation}

where

\begin{equation}
w_{c}(x) = \sinh^{-1}(\frac{1-x}{2\sqrt{x}})
\label{A6}
\end{equation}

(Eq. (A1.16) of ref. I) and hence

\begin{equation}
F(x,M) \sim 2(1-x) - 2\pi \sqrt{\frac{2(1-x^{2})}{xM}} e^{-2Mw_{c}(x)}
\label{A7}
\end{equation}
which exhibits the leading finite-size corrections, as required.


\setdec 0.00000000000
\begin{table}
\caption{DMRG estimates of the ground state energy density $ \omega_0/2Nx $, 
the ``2-particle'' gap $ \Delta_2/g $, and two order 
parameters, the mean field $ \Gamma^{\alpha} = \langle (L + \alpha) \rangle_0  
$ and axial density $ \Gamma^{5} = \langle i \bar{\psi} \gamma_5 \psi/g 
\rangle_0  $ as functions of $ l $, the number of basis states retained per 
block. These results are for periodic boundary conditions, at $ x = 100 $, 
$ m/g = 0.3 $, $ \theta = \pi $, and $ N = 256 $ sites. }
\label{tab:convpbcb0.5}
\begin{tabular}{ccccc}
$ l $  & $ \omega_0/2Nx $	& $ \Delta_{2}/g $ & 
$ \Gamma^{\alpha}  $ & $ \Gamma^{5}  $   \\
\hline
244  & -.31676292 & .26833 & .28068 & .29104 \\
324  & -.31677001 & .25687 & .27598 & .28147 \\
416  & -.31677263 & .25279 & .27417 & .27735 \\
555  & -.31677385 & .25054 & .27319 & .27528 \\
742  & -.31677428 & .24963 & .27273 & .27444 \\
932  & -.31677440 & .24933 & .27266 & .27423 \\ 
\end{tabular}
\end{table}
\begin{table}
\caption{As for Table \ref{tab:convpbcb0.5}, with open boundary conditions. }
\label{tab:convobcb0.5}
\begin{tabular}{ccccc}
$ l $  & $ \omega_0/2Nx $	& $\Delta_{2}/g $ & $ \Gamma^{\alpha} $ & 
$ \Gamma^{5}  $   \\
\hline
107   & -.31611009314180 & .19536745 & .256555 & -.257439 \\
162   & -.31611009382342 & .19535871 & .256594 & -.257479 \\
199   & -.31611009385738 & .19535765 & .256606 & -.257489 \\
244   & -.31611009386852 & .19535733 & .256782 & -.257658 \\
327   & -.31611009387277 & .19535723 & .256751 & -.257627 \\
396   & -.31611009387248 & .19535720 & .256734 & -.257604 \\ 
\end{tabular}
\end{table}
\begin{table}
\caption{Estimates for critical exponents $ 1/\nu $ and $ \beta/\nu $ as 
functions of the lattice spacing parameter $ x = 1/g^{2}a^{2}$. }
\label{tab:critexp}
\begin{tabular}{ccc}
$x$ & $ 1/\nu $ & $ \beta / \nu $\\
\hline
4.0 & 0.99(1) & 0.126(5) \\
4.938 & 1.00(2) & 0.125(5) \\
6.25 & 0.99(2) & 0.125(5) \\
8.163 & 0.99(2) & 0.125(6) \\
11.1 & 1.00(4) & 0.126(6) \\
16.0 & 0.99(3) & 0.126(6) \\
25.0 & 0.99(3) & 0.127(6)\\
44.4 & 0.97(4) & 0.123(6) \\
100.0 & 1.0(1) & 0.12(1) \\
\end{tabular}
\end{table}
\begin{table}
\caption{Our results for the loop energy $ \Delta_0/g $, 1-particle gap $ \Delta_1/g $, 2-particle gap $ \Delta_2/g $ at background field $ \theta = \pi $. We also quote our results for the order parameters $ \Gamma^{\alpha} = \langle (L + \alpha) \rangle_0  
$ and $ \Gamma^{5} = \langle i \bar{\psi} \gamma_5 \psi /g \rangle_0  $. }
\label{tab:bfieldresults}
\begin{tabular}{llllll}
$ m/g $ & $ \Delta_0/g $ & $ \Delta_1/g $  & $ \Delta_2/g $ & $ \Gamma^\alpha $ & $ \Gamma^5 $ \\
\hline
0.0	& 0.5643(2)  &$0.0 \pm 10^{-6}$& 0.57(1) &$0.0 \pm 10^{-4}$ &$0.0 \pm 10^{-4}$\\
0.05	& 0.4756(2)  &		   & 0.48(1) &   	 & 		\\
0.1 	& 0.3883(2)  &  $ 0.0 \pm 10^{-6} $ & 0.40(1) &$ 0.0 \pm 10^{-3} $&$0.0 \pm 10^{-2}	$\\
0.15	& 0.3020(5)  &		   & 0.30(2)   &  	 &		\\
0.2	& 0.2173(5)  & $ 0.0 \pm 10^{-5} $ & 0.23(4)   &  0.00(2) & 0.000(5) \\
0.25	& 0.134(2)   &		   & 0.16(4)   & 	&		\\
0.3	& 0.05(2)    & 	$ 0.0 \pm 10^{-2} $   & 0.03(7)   & 0.0(3)	 & 0.0(2)	\\
0.4	& $0.0 \pm 10^{-3}$ & 0.105(3)	   & 0.22(1)   & 0.376(1)  & 0.302(5)	\\
0.5	&  $0.0 \pm 10^{-4}$ & 0.246(3)	   & 0.49(1)   & 0.421(1)  & 0.270(5)	\\
0.6	&  $ 0.0 \pm 10^{-4} $ & 0.3764(6)   & 0.758(8)   & 0.4430(5) & 0.238(5)	\\
0.7	& $ 0.0 \pm 10^{-4} $& 0.5020(2)   & 1.006(4)   & 0.4566(5) & 0.211(5)	\\
0.8	& $0.0 \pm 10^{-5}$ & 0.6224(1)   & 1.249(4)   & 0.4657(5) & 0.189(5)	\\
1.0	& $0.0 \pm 10^{-5}$ & 0.8530(4)   & 1.711(4)   & 0.4769(5) & 0.155(5)	\\
\end{tabular}
\end{table}
\begin{table}
\caption{Comparison of bound-state energies for the ``vector'' state with previous works. The results of Sriganesh {\it et al.} and Crewther and Hamer were obtained through finite-lattice studies, while Eller {\it et al.}, Mo and Perry, and Kr\"{o}ger and Scheu used light-cone or related methods to obtain their results. 
}
\label{tab:vector}
\begin{tabular}{lllllll}
$ m/g $  & This work  & Sriganesh  & Crewther \&  
 &
Eller   & Mo \&  & Kr\"{o}ger \&   \\

    &       & {\it et al.} \cite{sriganesh00} & Hamer \cite{crewther80} & {\it et al.} \cite{eller87}  &  
 Perry \cite{mo93} & Scheu \cite{kroger98} \\
\hline
0     & 0.56419(4) & 0.563(1)  & 0.56(1)   &         &        & \\
0.125 & 0.53950(7) & 0.543(2)  & 0.54(1)   &  0.58 &    0.54& 0.528 \\
0.25  & 0.51918(5) & 0.519(4)  & 0.52(1)   &  0.53 &    0.52& 0.511 \\
0.5   & 0.48747(2) & 0.485(3)  & 0.50(1)   &  0.49 &    0.49& 0.489 \\
1     & 0.4444(1)  & 0.448(4)  & 0.46(1)   &  0.45 &    0.45& 0.445\\
2     & 0.398(1)   & 0.394(5)  & 0.413(5)  &  0.40 &    0.40& 0.394\\
4     & 0.340(1)   & 0.345(5)  & 0.358(5)  &  0.34 &    0.34& 0.339\\
8     & 0.287(8)   & 0.295(3)  & 0.299(5)  &  0.28 &    0.29& 0.285\\
16    & 0.238(5)   & 0.243(2)  & 0.245(5)  &  0.23 &    0.24& 0.235\\
32    & 0.194(5)   & 0.198(2)  & 0.197(5)  &  0.20 &    0.20& 0.191\\
\end{tabular}
\end{table}

\center
\widetext
\input psfig
\psfull
\begin{figure}
\centerline{\psfig{file=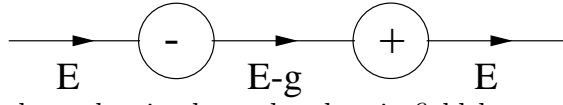,clip=}}
\caption{Creation of a charged pair alters the electric field by amount $ \pm g $ in the intervening space.}
\label{fig:pair}
\end{figure}
\begin{figure}
\centerline{\psfig{file=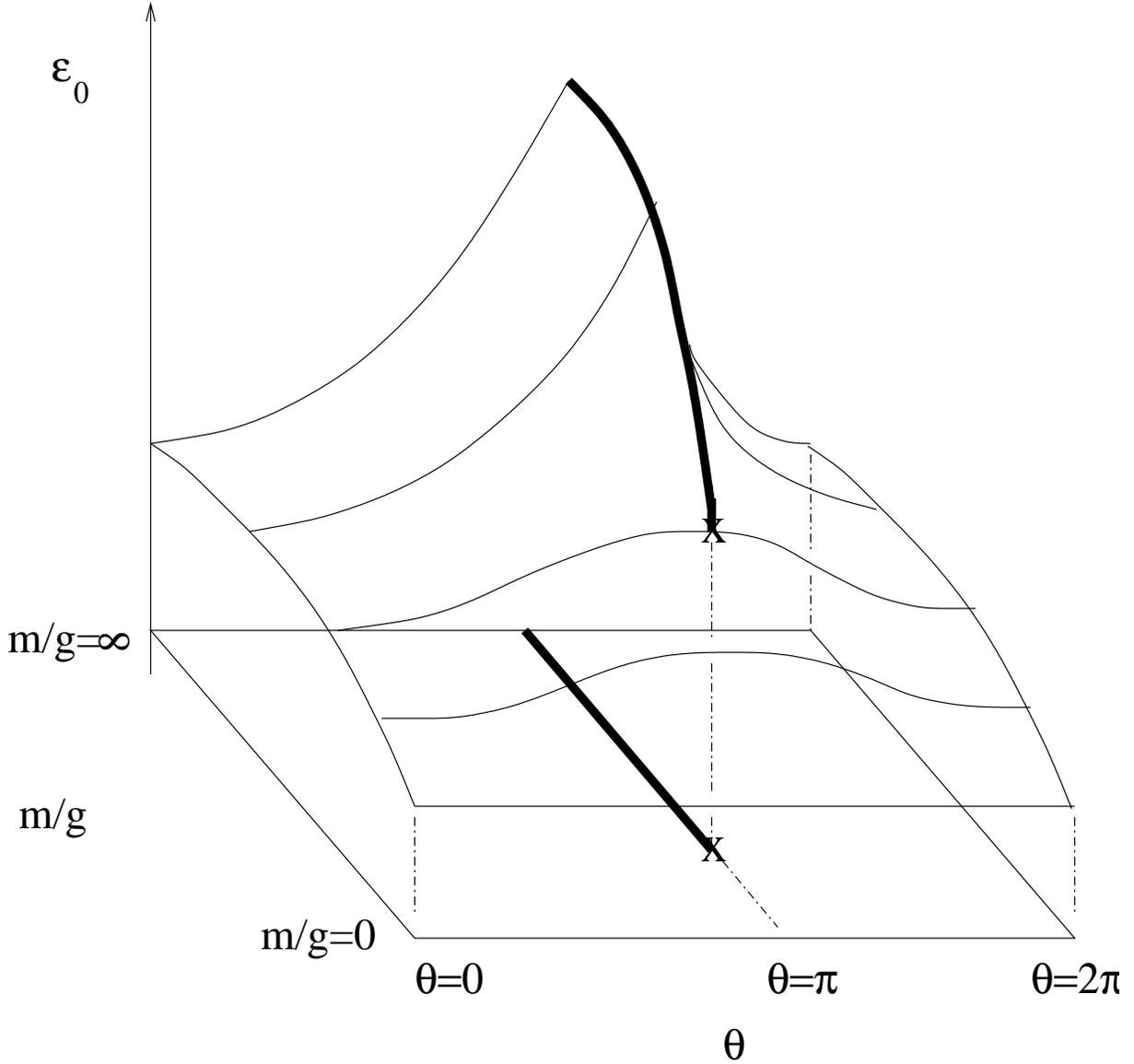,clip=}}
\caption{Schematic plot of the vacuum energy density as a function of $ m/g $ and $ \theta $. The heavy line marks the first-order transition line where the energy density has a cusp, terminating at the second order critical point $ (m/g)_c $, where the discontinuity in slope with $ \theta $ goes to zero.}
\label{fig:phase}
\end{figure}
\newpage
\begin{figure}
\centerline{\psfig{file=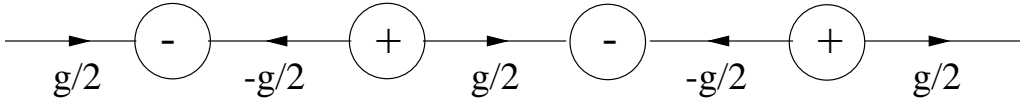,clip=}}
\caption{A configuration of ``half-asymptotic'' charged fermions at background field $ F = g/2 $.}
\label{fig:halfass}
\end{figure}
\begin{figure}
\centerline{\psfig{file=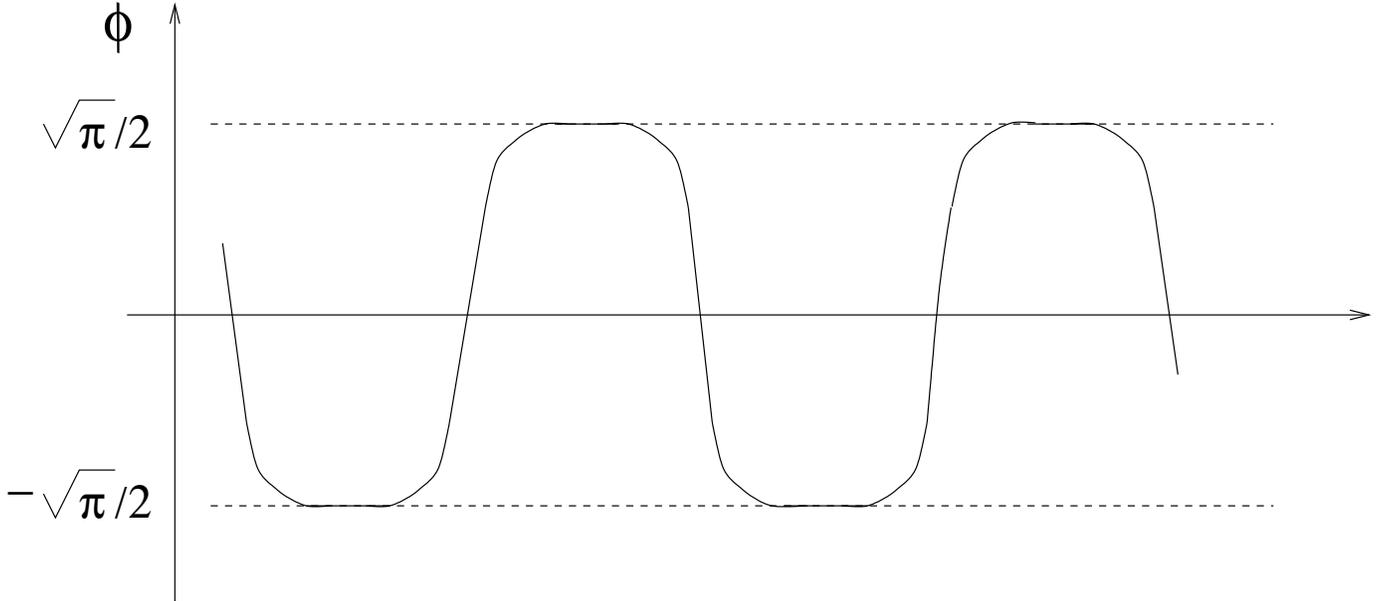,clip=}}
\caption{``Kinks'' and ``Antikinks'' are represented by transitions
between 
the ground state solutions $ \phi = \pm \frac{1}{2} \sqrt{\pi} $. Every
transition corresponds to either an quark or an antiquark, according to
the picture
given in Fig. \ref{fig:halfass}}   
\label{fig:kink}
\end{figure}
\begin{figure}
\centerline{\psfig{file=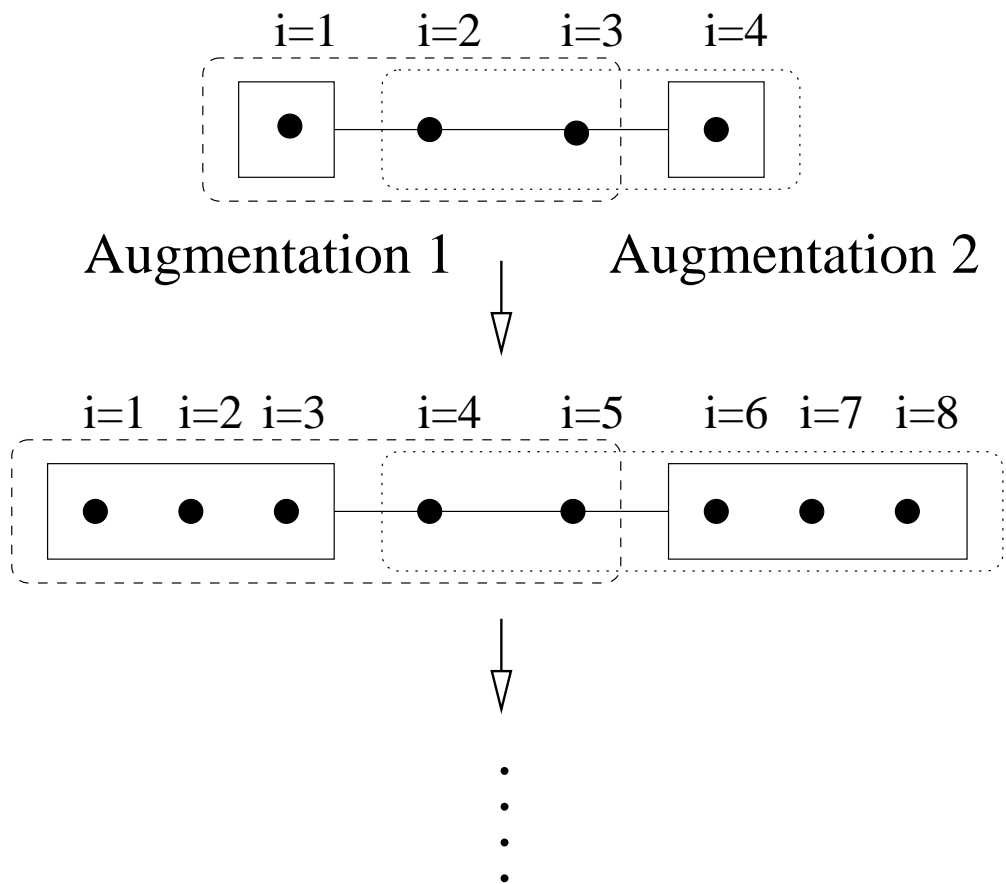,clip=}}
\caption{A DMRG iteration using open boundary conditions. Two augmentations are needed to allow for the difference between odd and even sites. }
\label{fig:dmrgobc}
\end{figure}
\begin{figure}
\centerline{\psfig{file=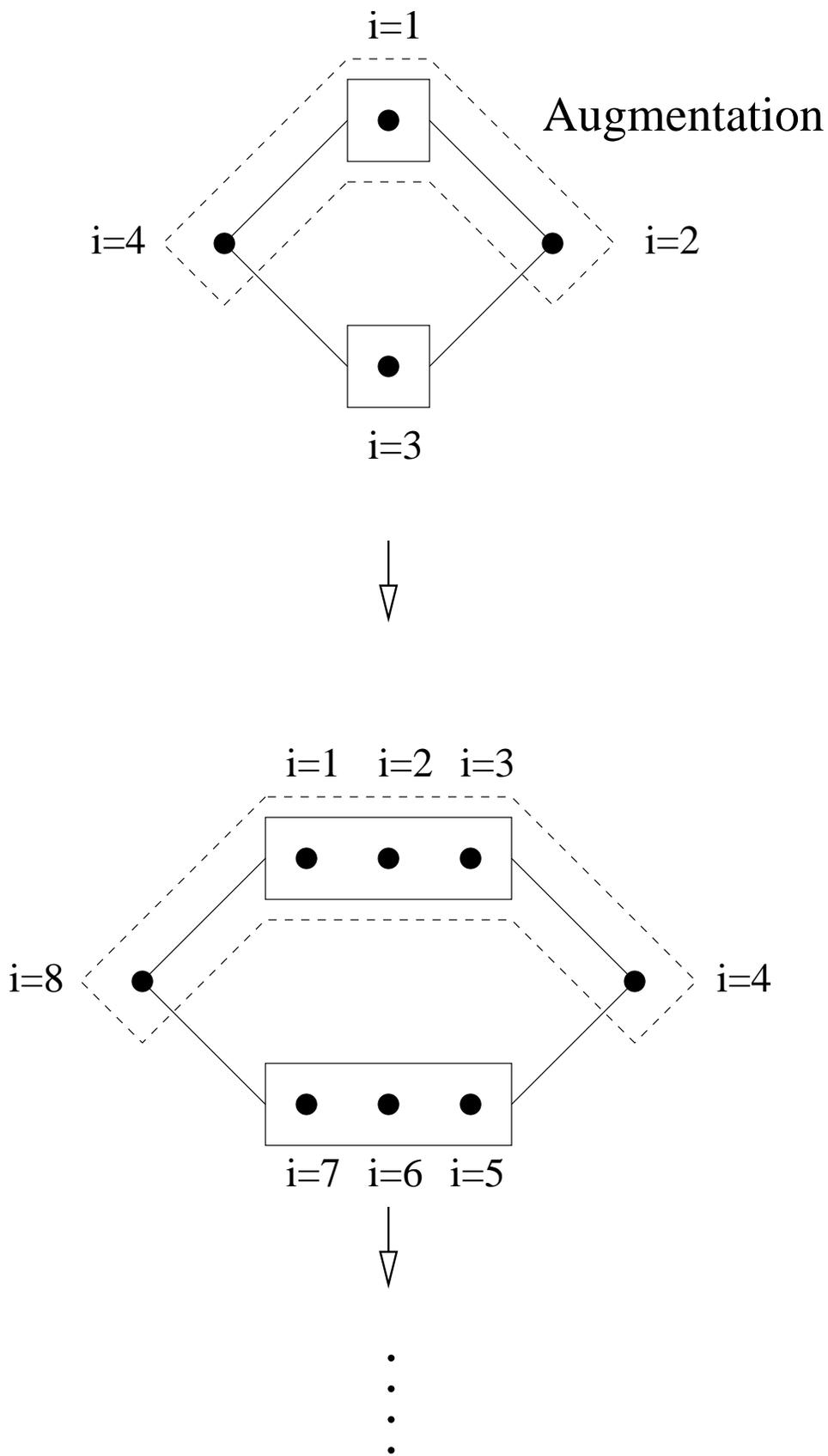,clip=}}
\caption{A DMRG iteration using periodic boundary conditions.}
\label{fig:dmrgpbc}
\end{figure}
\begin{figure}
\centerline{\psfig{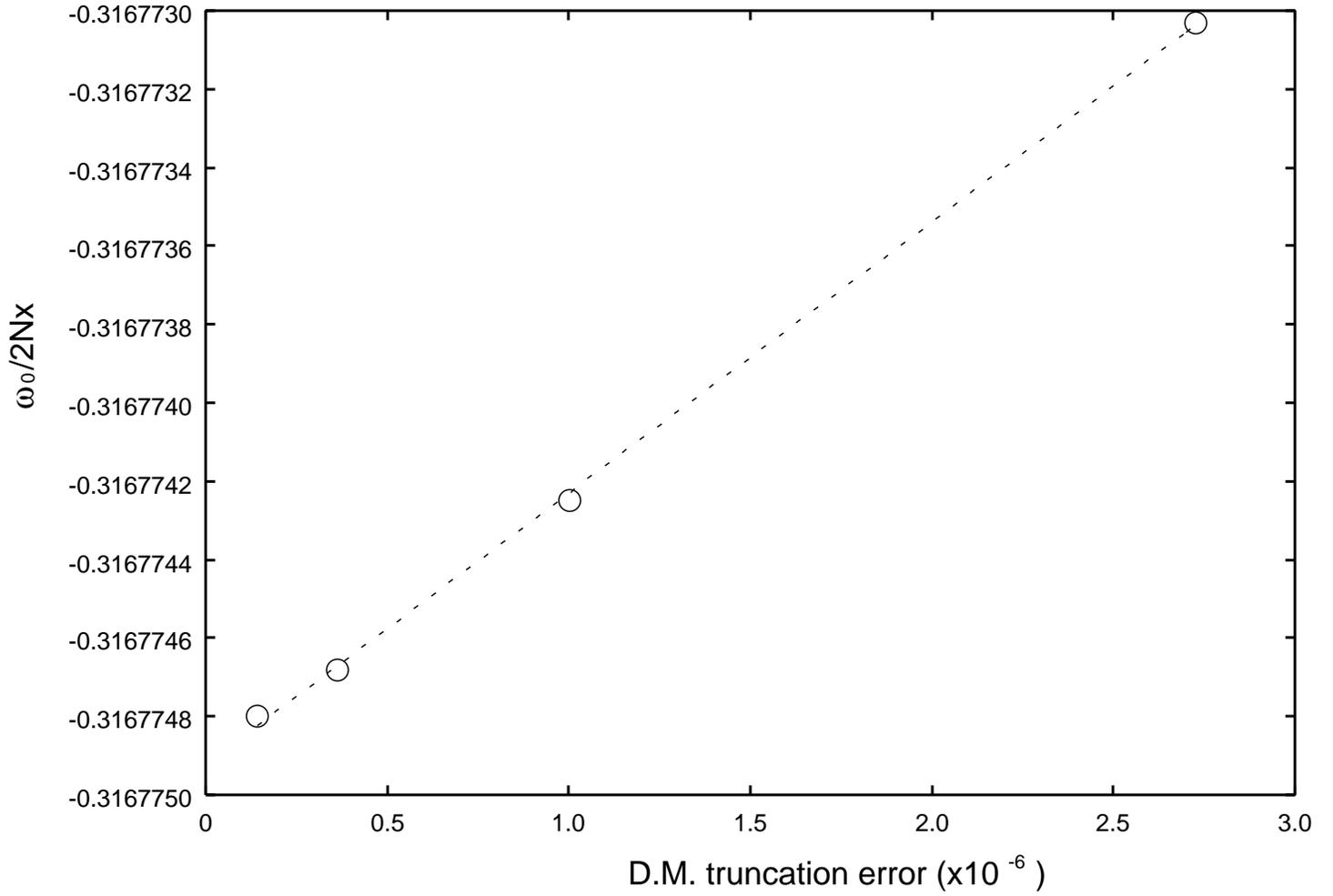}}
\caption{Dependence of the estimated ground state energy density $ \omega_0 / 2Nx $ on the density matrix truncation eigenvalues, for $ x = 100 $, $ m/g = 0.3 $, $ \theta = \pi $, using PBC. }
\label{fig:dmrgconvpbc}
\end{figure}
\begin{figure}
\centerline{\psfig{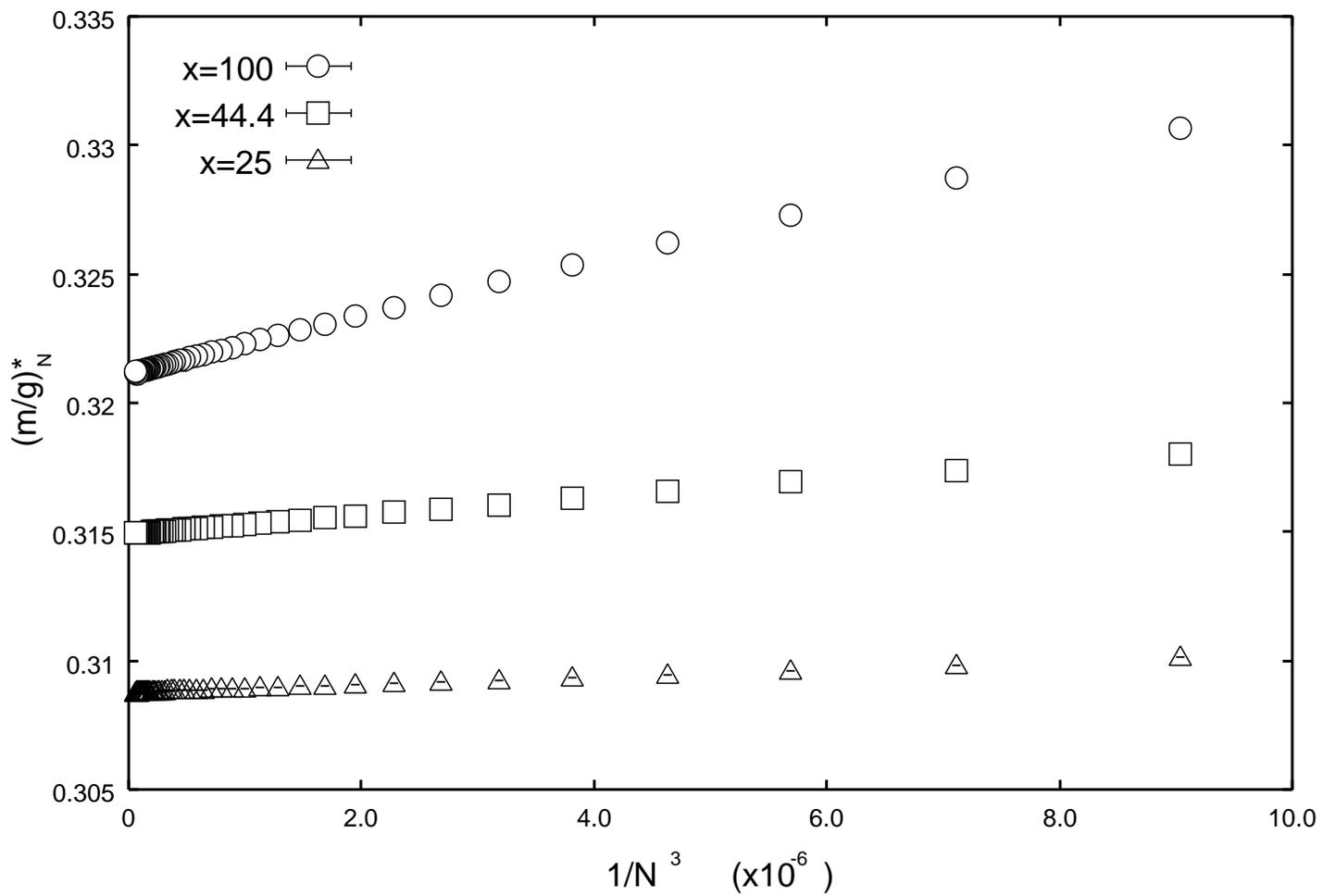}}
\caption{Pseudocritical points for various lattice spacings. }
\label{fig:pseudoplot}
\end{figure}
\begin{figure}
\centerline{\psfig{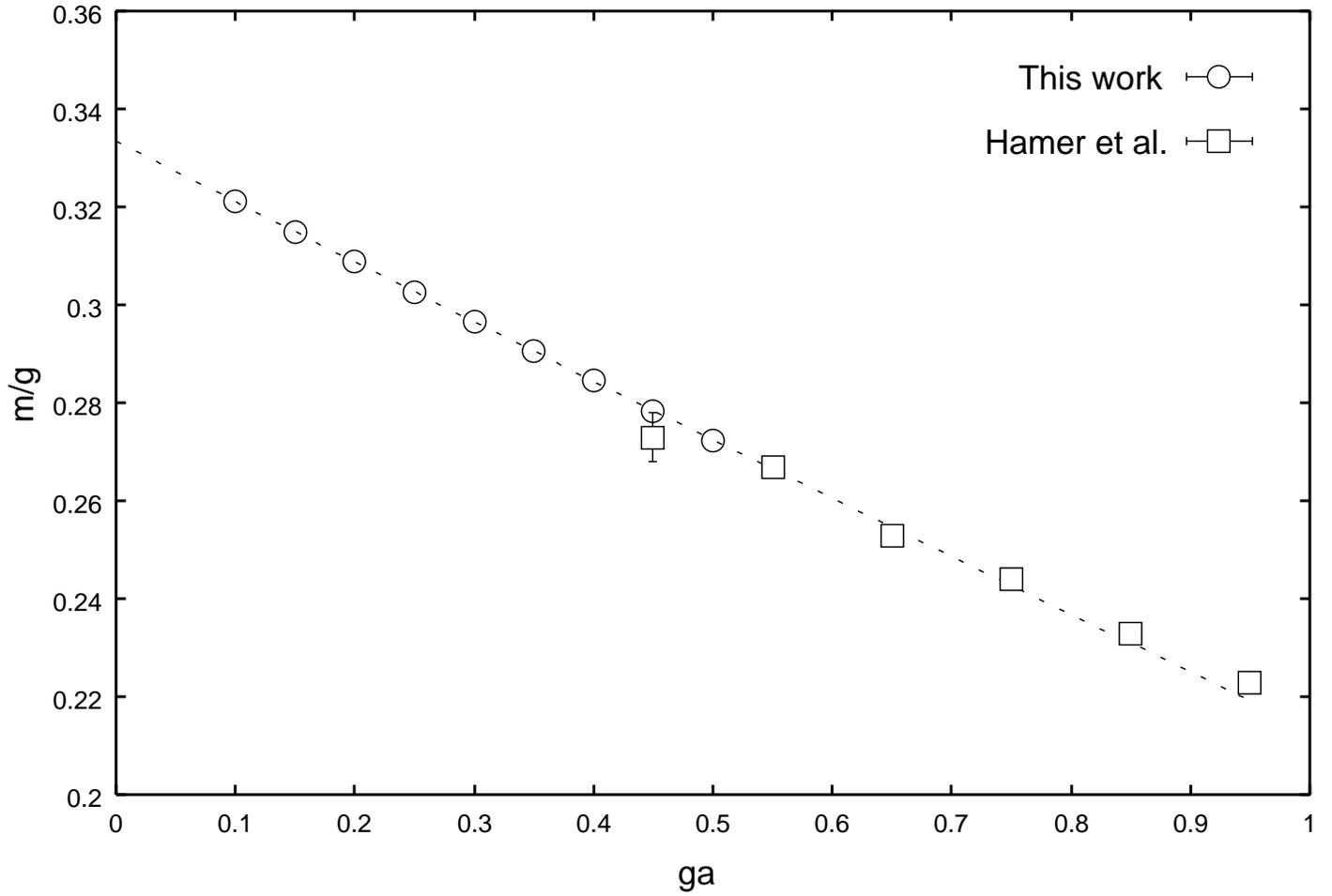}}
\caption{Critical line in the $ m/g $ versus $ 1/\sqrt{x} = ga $ plane. Open circles are our present estimates, and squares are the previous results of Hamer {\it et al.} \protect\cite{hamer82}, which are in good agreement. The dashed line is a quadratic fit to the data in $ ga $. }
\label{fig:finalplot}
\end{figure}
\begin{figure}
\centerline{\psfig{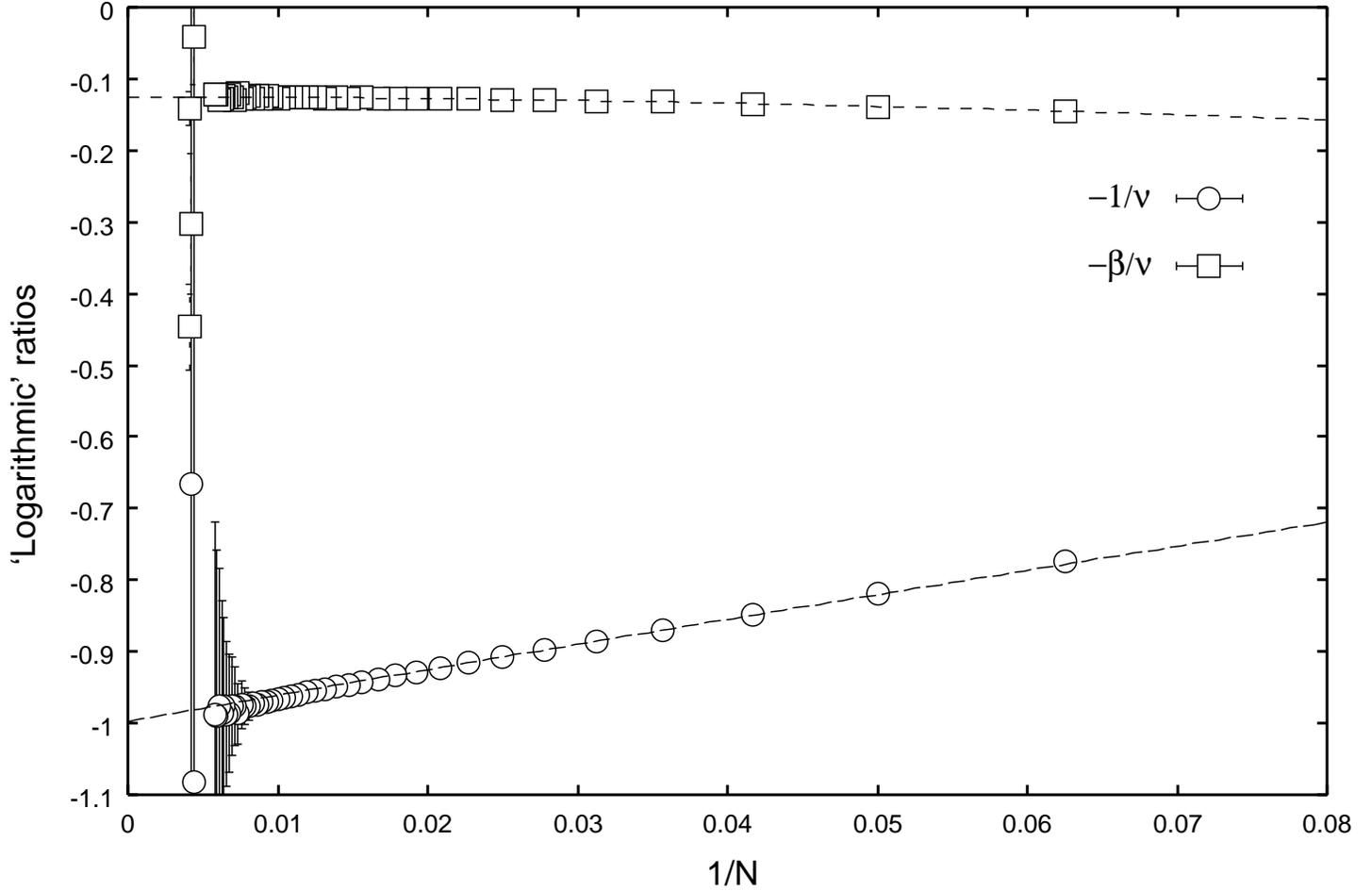}}
\caption{`Logarithmic' ratio estimates of critical indices $ - 1/\nu $ and 
$ - \beta / \nu $ for lattice spacing $ 1/\sqrt{x} = ga = 0.45 $. Quadratic 
fits in $ 1/N $ provide the bulk extrapolations. We estimate here $ 1/\nu = 
1.00(2) $ and $ \beta/\nu = 0.125(5) $. }
\label{fig:ratios}
\end{figure}
\begin{figure}
\centerline{\psfig{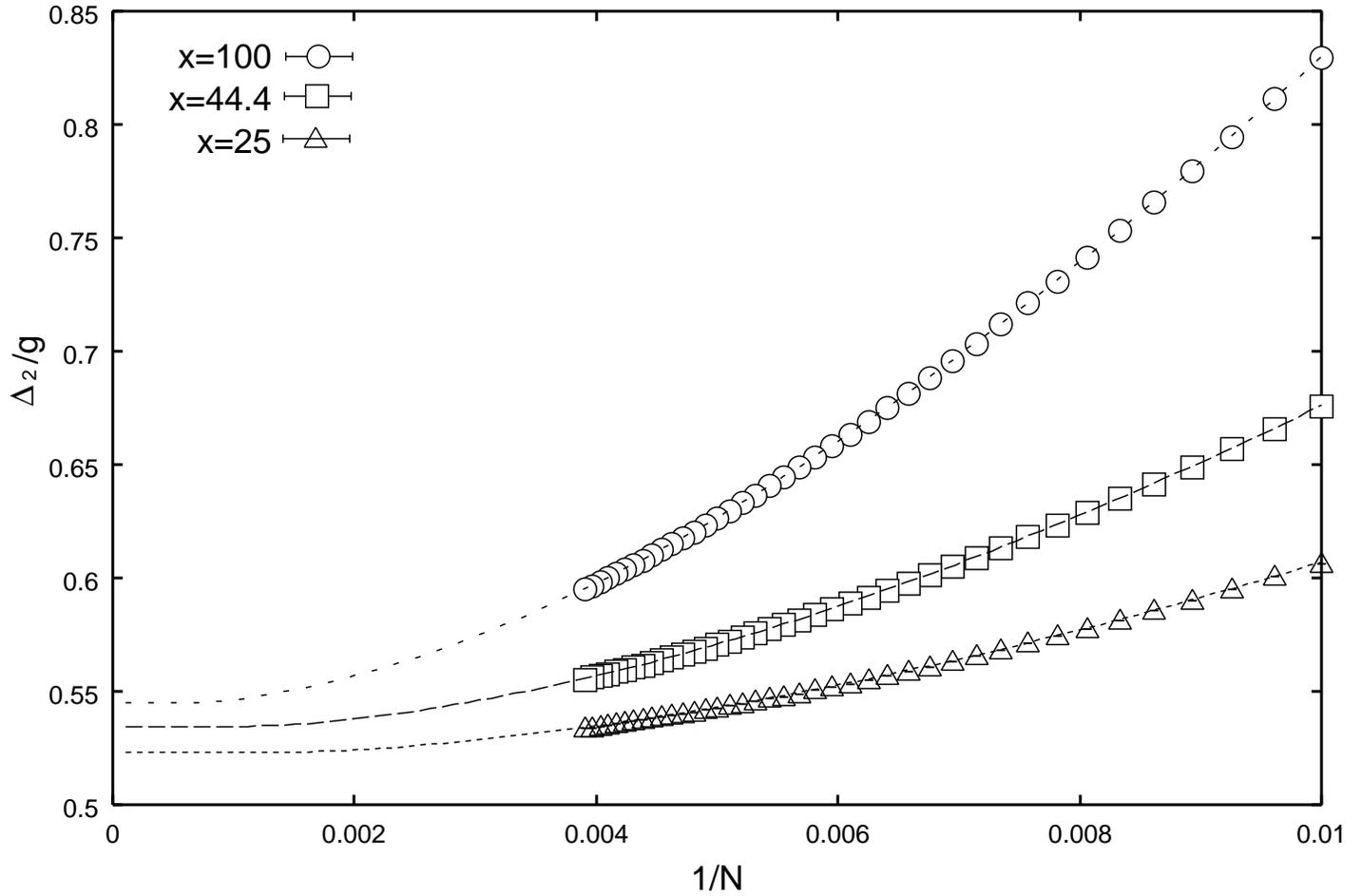}}
\caption{Bulk extrapolations for the 2-particle gap $ \Delta_2 /g $ for $ m/g = 0.0 $ and $ \theta = \pi $. Dashed lines are the fits to the data, according to (\ref{fitfunc})}
\label{fig:bulk2part}
\end{figure}
\begin{figure}
\centerline{\psfig{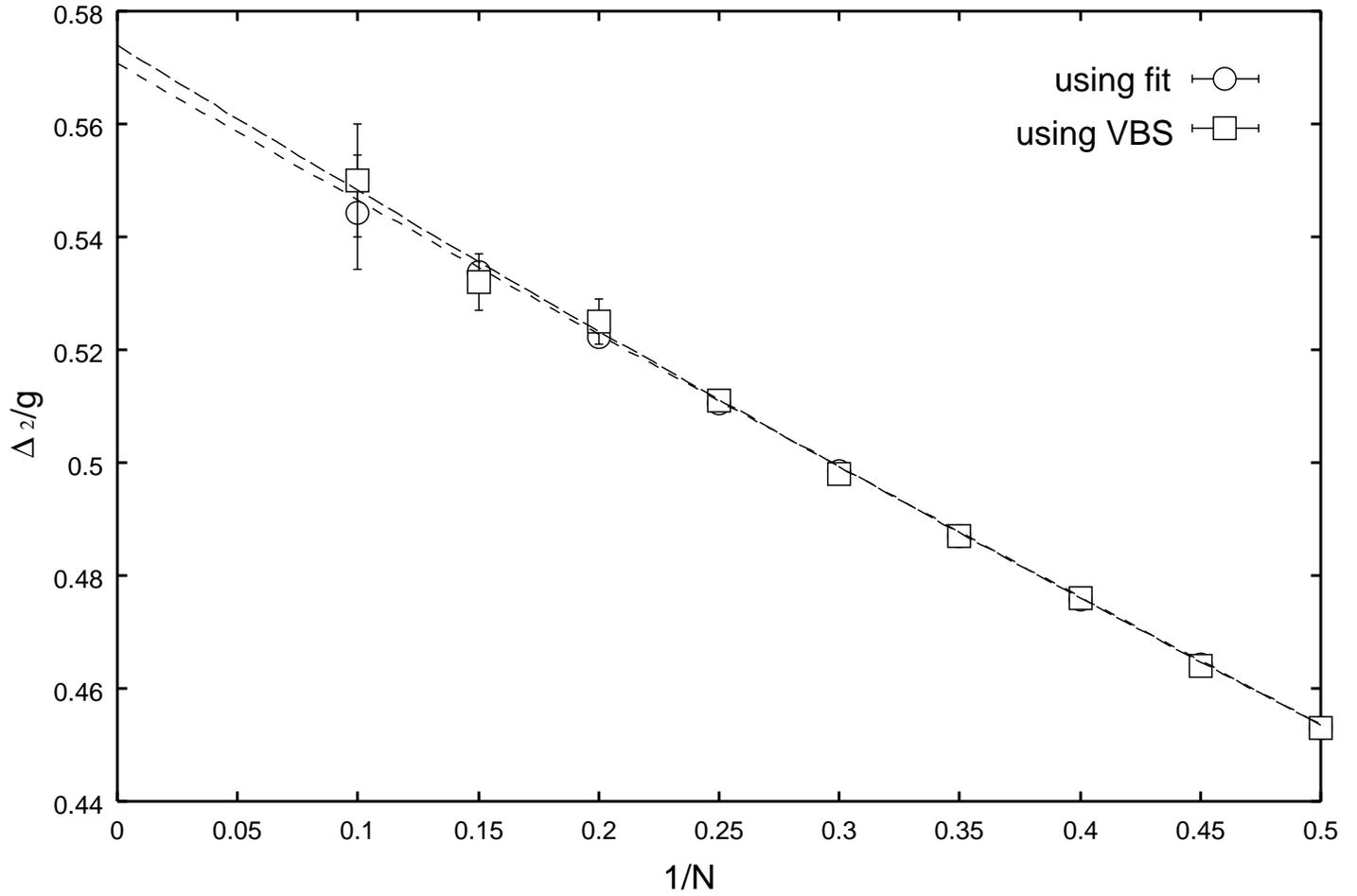}}
\caption{Continuum extrapolations for the 2-particle gap for $ m/g = 0.0 $ and $ \theta = \pi $. Data sets obtained through separate VBS and fit extrapolations. }
\label{fig:cont2part}
\end{figure}
\begin{figure}
\centerline{\psfig{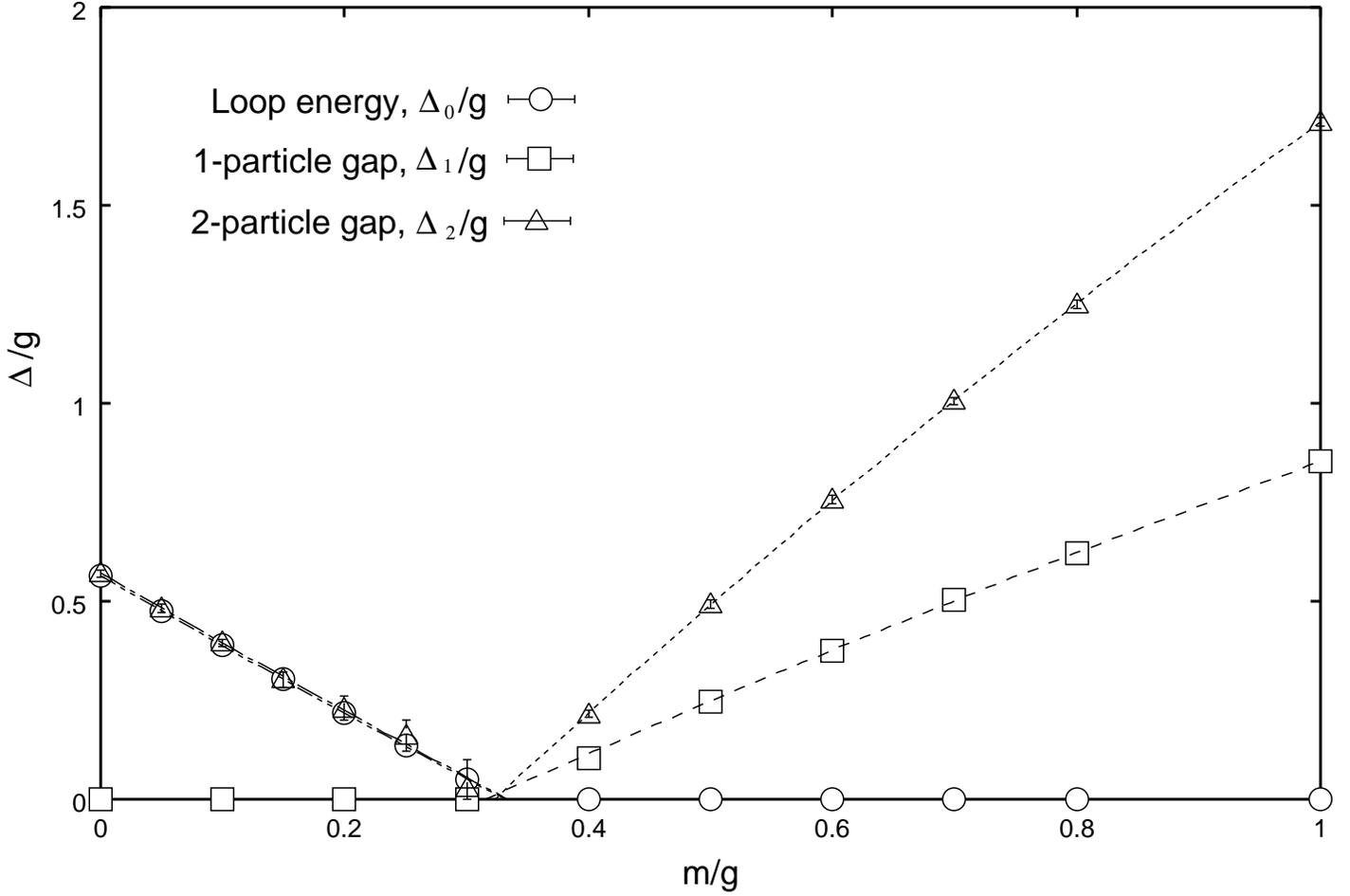}}
\caption{Final estimates for gaps in the 0-particle, 1-particle and
2-particle sectors at $ \theta = \pi $. Dashed lines are merely to guide the eye.}
\label{fig:mogland}
\end{figure}
\begin{figure}
\centerline{\psfig{file=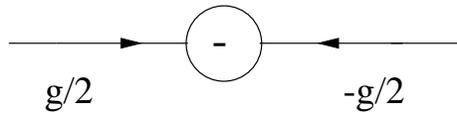,clip=}}
\caption{The electric fields at either end of the chain in the presence of a single particle are $ g/2, -g/2 $ respectively. This precludes periodic boundary conditions due to the mismatch of electric fields. }
\label{fig:halfasslone}
\end{figure}
\begin{figure}
\centerline{\psfig{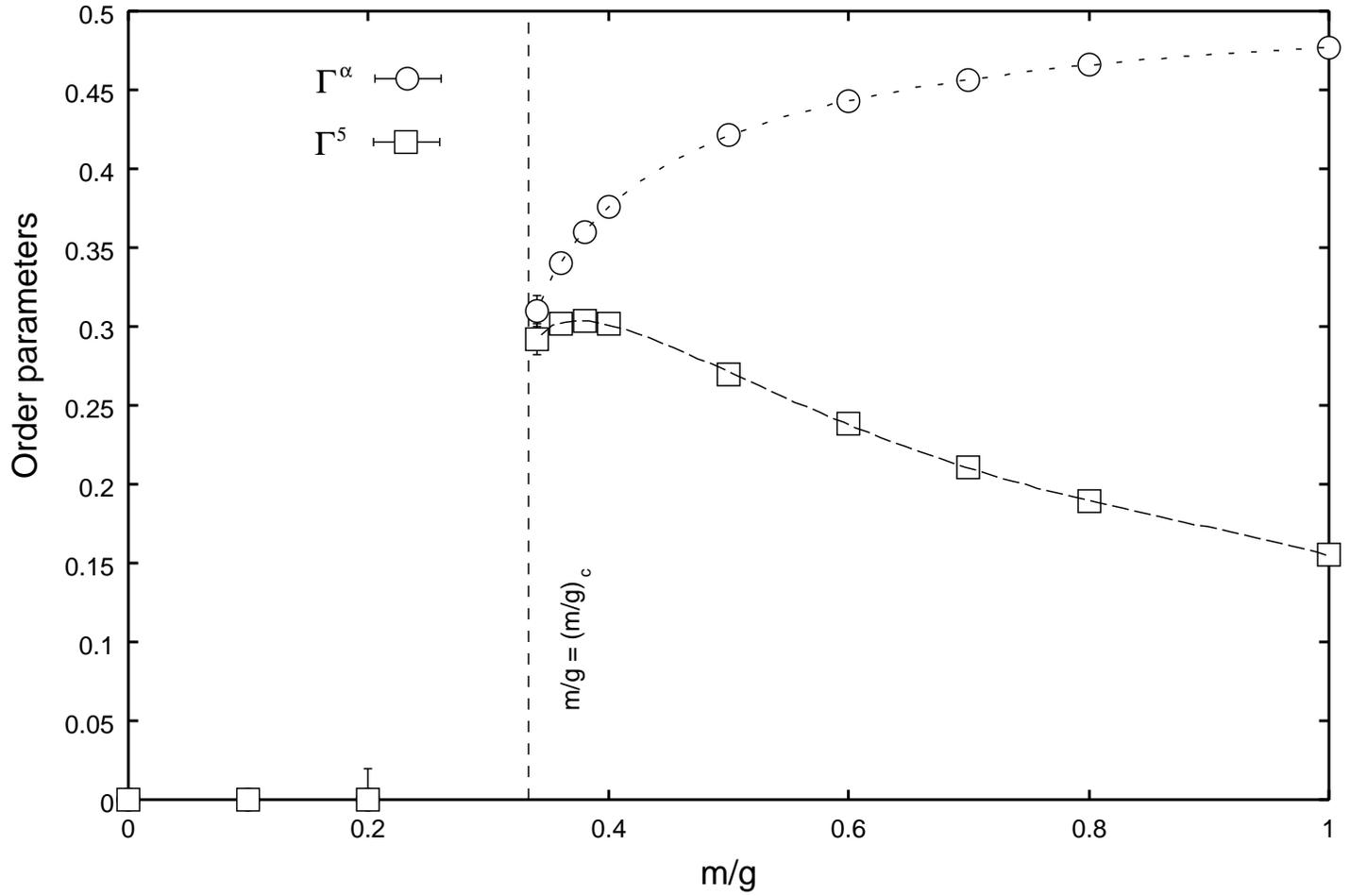}}
\caption{Order parameters  $ \Gamma^{\alpha} = \langle (L + \alpha) \rangle_0  
$, $ \Gamma^{5} = \langle i \bar{\psi} \gamma_5 \psi /g \rangle_0  $ near the critical region. Dashed lines are merely to guide the eye.}
\label{fig:moglandorder}
\end{figure}
\begin{figure}
\centerline{\psfig{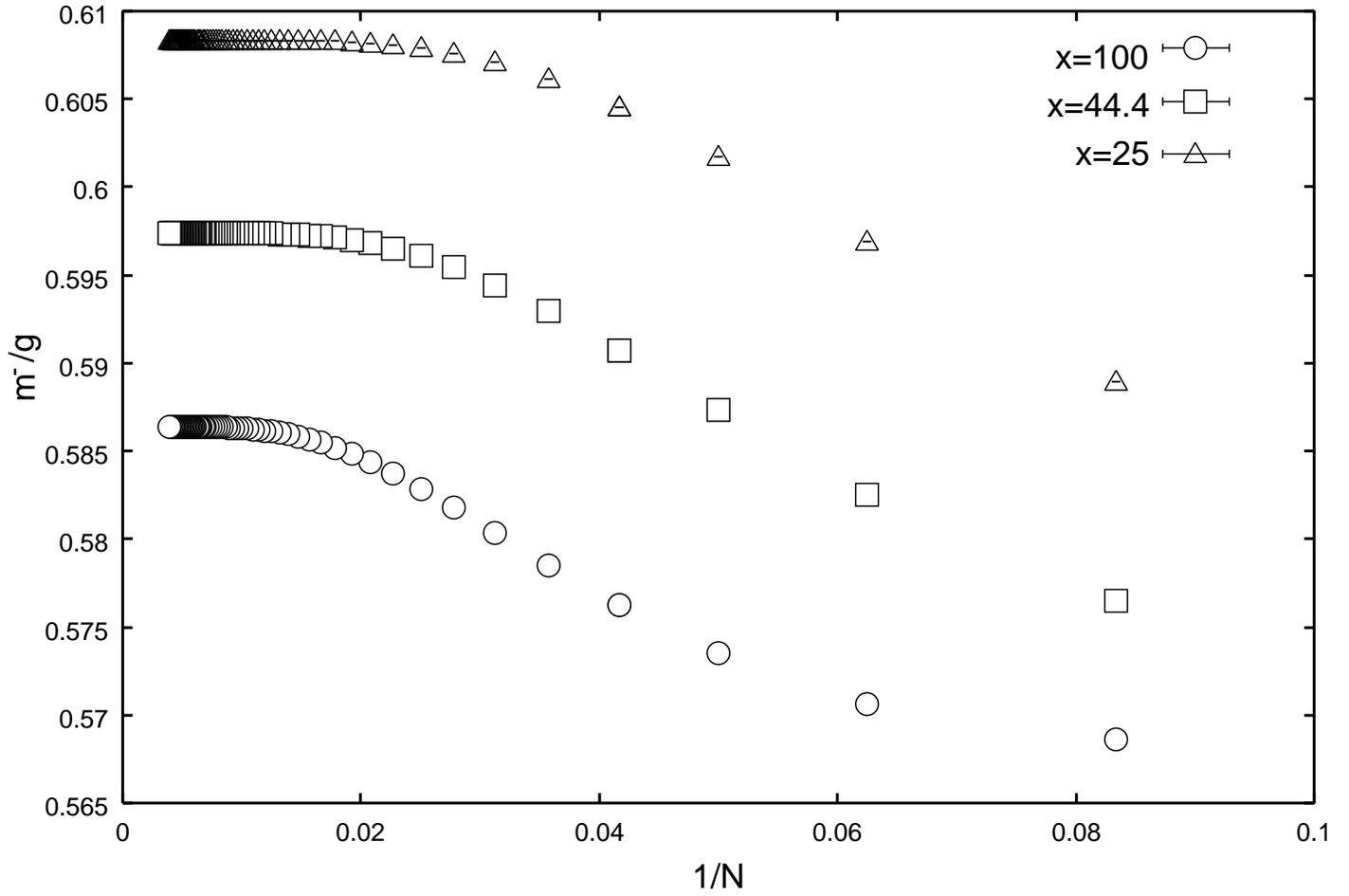}}
\caption{``Vector'' mass gaps $ m^- /g $ for $ m/g = 0 $, finite lattices $ N = 10$-$ 256 $, and various lattice spacings.}
\label{fig:vectorm0}
\end{figure}
\begin{figure}
\centerline{\psfig{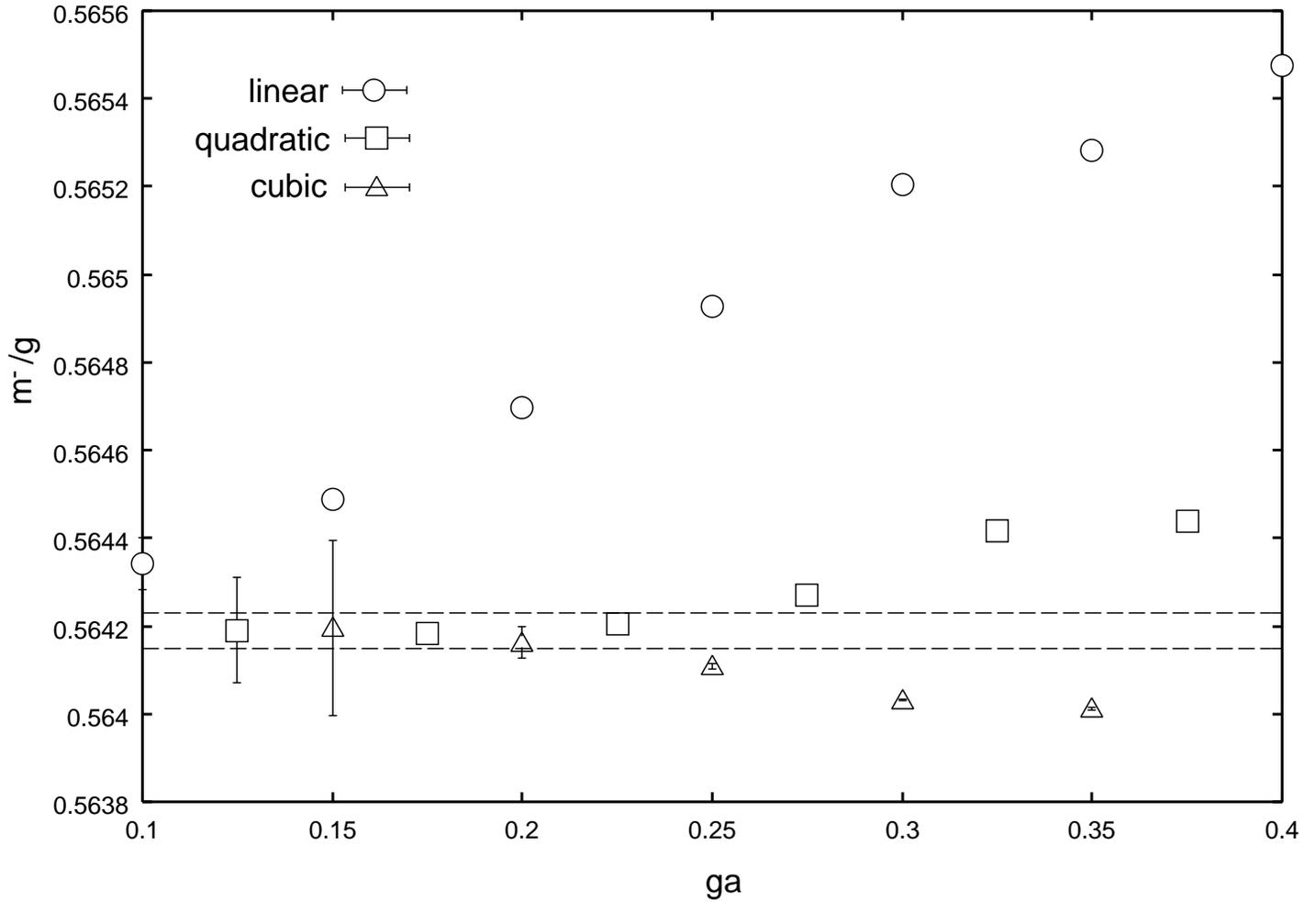}}
\caption{An example of our extrapolation procedure for the ``vector'' mass 
gaps at $ m/g = 0 $. Circles, squares and triangles show linear, quadratic and 
cubic extrapolants respectively. Dashed lines show the upper and lower bounds 
for our final estimate. Here we estimate $ m^-/g = 0.56419(4) $.  }
\label{fig:linquadcubm0}
\end{figure}
\begin{figure}
\centerline{\psfig{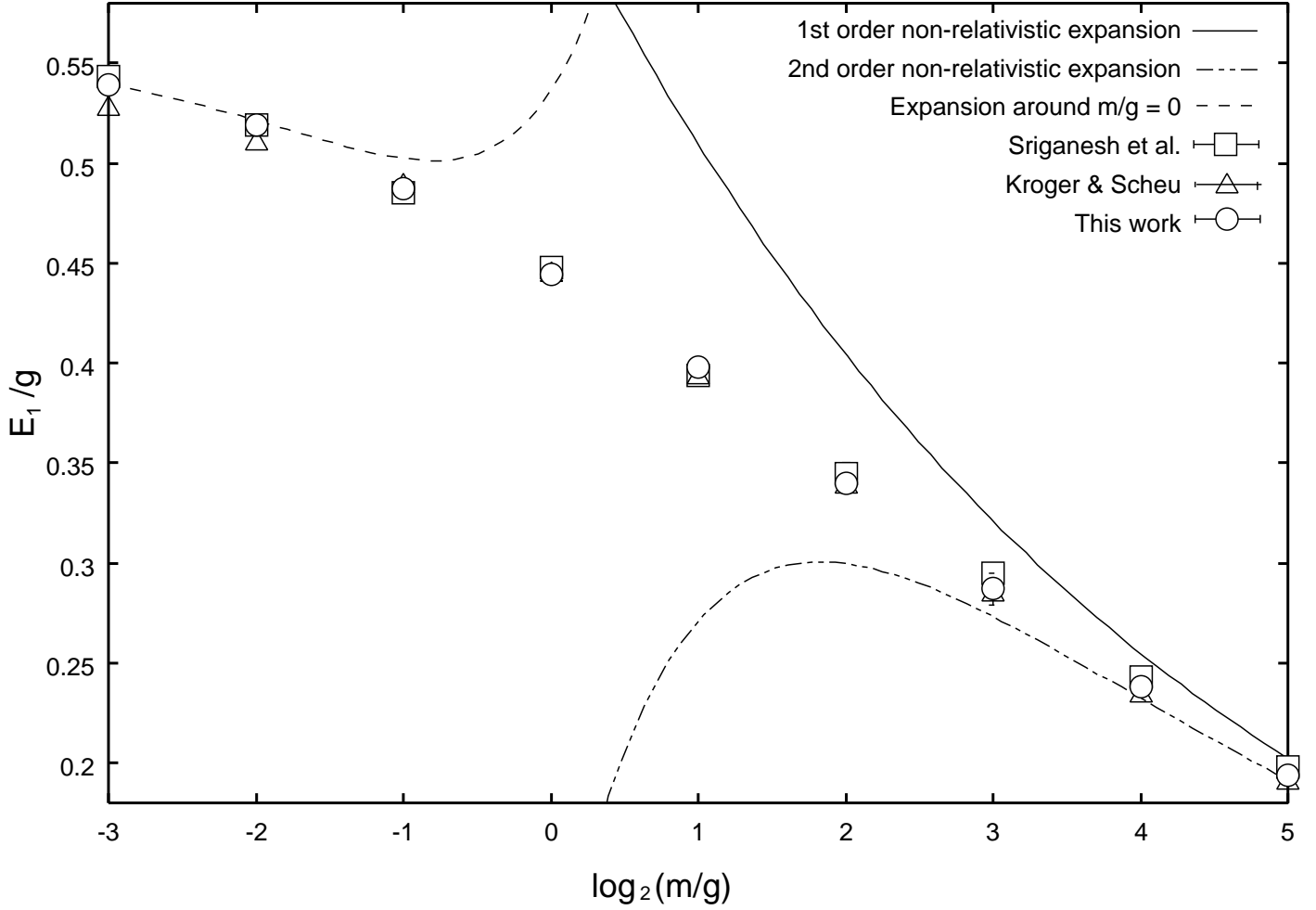}}
\caption{Comparison of our results for the ``vector'' state binding energies $ E_1/g $ with other works. Squares mark the results of Sriganesh {\it et al.} \protect\cite{sriganesh00} and triangles the results of Kr\"{o}ger and Scheu \protect\cite{kroger98}. 
The results of Vary, Fields and Pirner \protect\cite{vary96} and Adam \protect\cite{adam96} were used for the expansion around $ m/g = 0 $, while the non-relativistic expansions were done in the works of Hamer \protect\cite{hamer77} and Sriganesh {\it et al.} \protect\cite{sriganesh00}. }
\label{fig:moglandvector}
\end{figure}


\begin{references}

\bibitem{schwinger62} J. Schwinger, Phys. Rev. {\bf 128}, 2425 (1962).

\bibitem{lowenstein71} J. Lowenstein and J. Swieca, Ann. of Phys. {\bf 68}, 172 (1971).

\bibitem{coleman75} S. Coleman, R. Jackiw, and L. Susskind,
   Ann. of Phys. {\bf 93}, 267 (1975).

\bibitem{casher74} A. Casher, J. Kogut, and L. Susskind, Phys. Rev. D {\bf 10}, 732 (1974).
\bibitem{casher75}A. Casher, J. Kogut and L. Susskind, Ann. Phys. (N.Y.) {\bf 93}, 267(1975).

\bibitem{coleman76} S. Coleman, Ann. of Phys. {\bf 101}, 239 (1976).

\bibitem{creutz95} M. Creutz, Nucl. Phys. Proc. Suppl. {\bf 42}, 56 (1995).

\bibitem{narayanan95}
R. Narayanan, H. Neuberger and P. Vranis, Phys. Letts. {\bf
B353}, 507 (1995).

\bibitem{gattringer96}
C. Gattringer, Phys. Rev. {\bf D53}, 5090 (1996).

\bibitem{kiskis00}
J. Kiskis and R. Narayanan, Phys. Rev. {\bf D62}, 054501 (2000).

\bibitem{white92} S.R. White, Phys. Rev. Lett. {\bf 69}, 2863 (1992); Phys. Rev. B {\bf 48}, 10345 (1993).

\bibitem{gehring97}
G.A. Gehring, R.J. Bursill and T. Xiang, Acta Phys. Pol. {\bf 91}, 105 
(1997).

\bibitem{hamer82} C.J. Hamer, J. Kogut, D.P. Crewther, and M.M. Mazzolini, Nucl. Phys. {\bf B208}, 413 (1982).

\bibitem{burden88} C.J. Burden and C.J. Hamer, Phys. Rev. D {\bf 37}, 479 (1988), Appendix.  

\bibitem{banks76} T. Banks, L. Susskind, and J. Kogut, Phys. Rev. D {\bf 13}, 1043 (1976).

\bibitem{carroll76} A. Carroll, J. Kogut, D.K. Sinclair, and L. Susskind,
Phys. Rev. D {\bf 13}, 2270 (1976).

\bibitem{berruto98} F. Berruto, G. Grignani, G. W. Semenoff and P. Sodano, Phys. Rev. {\bf 57}, 5070 (1998)  

\bibitem{hamer97}C.J. Hamer, Zheng Weihong and J. Oitmaa, Phys. Rev. {\bf
D56}, 55 (1997).

\bibitem{crewther80} D.P. Crewther and C.J. Hamer, Nucl. Phys. B {\bf 170}, 353 (1980).

\bibitem{irving83}A.C. Irving and A. Thomas, Nucl. Phys. B {\bf 215}, 23
(1983).

\bibitem{sriganesh00} P. Sriganesh, C.J. Hamer and R.J. Bursill, Phys. Rev. D {\bf 62}, 034508 (2000). 

\bibitem{martin82}
O. Martin and S. Otto, Nucl. Phys. {\bf B203}, 297 (1982)

\bibitem{schiller82}
A.J. Schiller and J. Ranft, Nucl. Phys. {\bf B225}, 204 (1983)

\bibitem{carson86}
S.R. Carson and R.D. Kenway, Ann. Phys. {\bf 166}, 364 (1986).

\bibitem{baillie87}
C.F. Baillie, Nucl. Phys. {\bf B283}, 217 (1987).

\bibitem{azcoiti94}
V. Azcoiti, G. Di Carlo, A. Galante, A.F. Grillo and V. Laliena, Phys.
Rev. {\bf D50}, 6994 (1994).

\bibitem{eller87} T. Eller, H.C. Pauli, and S.J. Brodsky, Phys. Rev. D {\bf 35}, 1493 (1987).

\bibitem{bergknoff77}H. Bergknoff, Nucl. Phys. B {\bf 122}, 215(1977).

\bibitem{mo93} Y. Mo and R.J. Perry, J. Comput. Phys. {\bf 108}, 159 (1993).

\bibitem{kroger98}H. Kr\"{o}ger and N. Scheu, Phys. Lett. B {\bf 429}, 58 (1998).

\bibitem{melnikov00}
K. Melnikov and M. Weinstein, Phys. Rev. {\bf D62}, 094504 (2000).

\bibitem{fang01}
X-Y. Fang, D. Sch{\" u}tte, V. Wethkamp and A. Wichmann, Phys. Rev. {\bf
D64}, 014501 (2001).

\bibitem{vary96}J.P. Vary, T.J. Fields and H.J. Pirner, Phys. Rev. D {\bf 53}, 7231(1996).

\bibitem{adam96}C. Adam, Phys. Lett. B {\bf 382}, 383(1996); Ann. Phys.

\bibitem{harada95} K. Harada, A. Okazaki and M. Taniguchi, Phys. Rev. D {\bf
52}, 2429 (1995);
K. Harada, T. Heinzl and Christian Stern, Phys. Rev. {\bf D57},
2460 (1998)

\bibitem{hamer77}C.J. Hamer, Nucl. Phys. B {\bf 121}, 159(1977).

\bibitem{mandelstam75} S. Mandelstam, Phys. Rev. D {\bf 11}, 3026 (1975).

\bibitem{martindelgado99}
M.A. Martin-Delgado and G. Sierra, Phys. Rev. Letts. {\bf 83}, 1514
(1999).

\bibitem{kogut75} J. Kogut and L. Susskind, Phys. Rev. D {\bf 11},
395 (1975).

\bibitem{yang}
C.N. Yang, Phys. Rev. {\bf 85}, 808 (1952); K. Uzelac and R. Jullien, J.
Phys. {\bf A14}, L151 (1981); C.J. Hamer, J. Phys. {\bf A15}, L675
(1982).

\bibitem{barber83} M.N. Barber, in {\it Phase Transitions and Critical Phenomena}, vol. 8, edited by C. Domb and J. Lebowitz (Academic, New York, 1983).

\bibitem{hamer81} C.J. Hamer and M.N. Barber, J. Phys. A {\bf 14}, 241 (1981).

\bibitem{vandenbroeck79} J.M. Vanden Broeck and L.W. Schwartz, SIAM (Soc. Ind. Appl. Math.) J. Math. Anal. {\bf 10}, 658 (1979); M.N. Barber and C.J. Hamer, J. Aust. Math. Soc. B, Appl. Math. {\bf 23}, 229 (1980). 


\bibitem{fradkin} E. Fradkin and L. Susskind, Phys. Rev. {\bf D17}, 2637 (1978).

\bibitem{schultz66} T. Schultz, D. Mattis and E. Lieb, Rev. Mod. Phys.
{\bf 36}, 856 (1964).






\end{references}
\end{document}